\documentclass[Transaction]{IEEEtran}
\usepackage{amsfonts,subfigure,multicol,color,verbatim,
graphicx,cite,epsfig,amssymb,amsmath,cases,bm,algorithm,
algorithmic,xcolor,multirow}

\newtheorem{prob}{Problem}

\begin{document}
\title{Secrecy-Optimized Resource Allocation for Device-to-Device Communication Underlaying Heterogeneous Networks}

\author{\normalsize
Kecheng~Zhang, Mugen~Peng,~\IEEEmembership{Senior Member,~IEEE}, Ping~Zhang,~\IEEEmembership{Senior Member,~IEEE}, and Xuelong~Li,~\IEEEmembership{Fellow,~IEEE}
\thanks{Copyright (c) 2015 IEEE. Personal use of this material is permitted. However, permission to use this material for any other purposes must be obtained from the IEEE by sending a request to pubs-permissions@ieee.org.}
\thanks{Kecheng~Zhang (e-mail: {\tt buptzkc@163.com}), Mugen~Peng (e-mail: {\tt pmg@bupt.edu.cn}), and Ping~Zhang (e-mail: {\tt pzhang@bupt.edu.cn}) are with the Key Laboratory of Universal
Wireless Communications for Ministry of Education, Beijing
University of Posts and Telecommunications (BUPT), China. Xuelong Li (e-mail: {\tt xuelong\_li@opt.ac.cn}) is with the Center for OPTical IMagery Analysis and Learning (OPTIMAL), State Key Laboratory of Transient Optics and Photonics, Xi'an Institute of Optics and Precision Mechanics, Chinese Academy of Sciences, Xi'an 710119, China.} }

\maketitle

\begin{abstract}
Device-to-device (D2D) communications recently have attracted much attention for its potential capability to improve spectral efficiency underlaying the existing heterogeneous networks (HetNets). Due to no sophisticated control, D2D user equipments (DUEs) themselves cannot resist eavesdropping or security attacks. It is urgent to maximize the secure capacity for both cellular users and DUEs. This paper formulates the radio resource allocation problem to maximize the secure capacity of DUEs for the D2D communication underlaying HetNets which consist of high power nodes and low power nodes. The optimization objective function with transmit bit rate and power constraints, which is non-convex and hard to be directly derived, is firstly transformed into matrix form. Then the equivalent convex form of the optimization problem is derived according to the Perron-Frobenius theory. A heuristic iterative algorithm based on the proximal theory is proposed to solve this equivalent convex problem through evaluating the proximal operator of Lagrange function. Numerical results show that the proposed radio resource allocation solution significantly improves the secure capacity with a fast convergence speed.
\end{abstract}

\section{Introduction}
Heterogeneous networks (HetNets), which are expected to significantly improve capacity and extend coverage, have been widely applied to affording the explosive growth of data traffic\textcolor[rgb]{1.00,0.00,0.00}{\cite{n1}}. A HetNet uses a mixture of high power nodes (HPNs) (e.g., macro or micro base stations (BSs)) and low power nodes (LPNs) (e.g., picocell BSs, femtocell BSs, small cell BSs, wireless relay, or distributed antennas) in the same coverage. HPNs are deployed to provide seamless connectivity and guarantee the basic quality of service (QoS) requirements of user equipments (UEs). Since the transmit power of HPNs is sufficiently high to achieve seamless coverage, all UEs often access the HPN to obtain the control signalling of the whole HetNet, which fulfills the decouple of control and user planes\textcolor[rgb]{1.00,0.00,0.00}{\cite{n2}}. In some hot spots with huge value of packet traffic, a UE prefers to access LPNs rather than HPNs because LPNs can provide much higher capacity than the HPN does. Therefore, the dense deployment of LPNs increases spectral efficiency (SE) and is one of the key factors for boosting the capacity of the future network\textcolor[rgb]{1.00,0.00,0.00}{\cite{n3}}. However, although the HetNet is a good alternative to improve SE with seamless coverage, the inter-tier interference between HPNs and LPNs is severe, and the backhaul is often constrained because UEs have to connect with the core network with a high capacity requirement\textcolor[rgb]{1.00,0.00,0.00}{\cite{Hetnet_peng}}. In order to alleviate the heavy load on backhaul, several alternative approaches have been presented. In particular, a hierarchical cloud computing architecture is proposed in \textcolor[rgb]{1.00,0.00,0.00}{\cite{review_add}} through adding a mobile dynamic cloud.

To alleviate the heavy burden on the backhaul and improve SE in HetNets, device-to-device (D2D) communication which is defined as direct communication between UEs without passing through BSs has been proposed in the $3^{rd}$ generation partnership project long term evolution-advanced\textcolor[rgb]{1.00,0.00,0.00}{\cite{1}}. With more traffic being transmitted by D2D communication, information exchange between UEs and core network is abated\textcolor[rgb]{1.00,0.00,0.00}{\cite{n4}}. Recently, D2D communication has attracted much attention with the ever increasing demand for the local content exchange between two nearby UEs. In D2D underlaying HetNets, radio resource is typically reused between the direct D2D link and the cellular air access link. As a result, the inter-tier interference is critical to make D2D communication rollout. Many works have been done to improve SE for the D2D communication underlaying HetNets. In\textcolor[rgb]{1.00,0.00,0.00}{\cite{2}}, the authors show that D2D communications can effectively improve the total throughput without generating harmful interference to HetNets with proper managment. In \textcolor[rgb]{1.00,0.00,0.00}{\cite{3}}, the harmful inter-system interference problems for D2D underlaying cellular networks are addressed, in which two interference avoidance mechanisms are applied to suppress the inter-tier interference between cellular and D2D links. Other recent researches, e.g.,\textcolor[rgb]{1.00,0.00,0.00}{\cite{4,5,6}}, also have verified that the D2D communication underlaying HetNets have several advantages, including the high spectral utilization, the low energy consumption, and incurring new packet services. To improve SE, the resource allocation should be optimized for D2D communication underlaying HetNets, which is a non-convex optimization problem. The non-convex optimization problem has been extensively researched, in which the Lagrangian dual method has been widely used\textcolor[rgb]{1.00,0.00,0.00}{\cite{r1,r2}}. However, it is assumed that the interference from cellular users to D2D users is constant in\textcolor[rgb]{1.00,0.00,0.00}{\cite{r1}} and the inter-tier interference between cellular users and D2D users is replaced by the maximum allowed noise in\textcolor[rgb]{1.00,0.00,0.00}{\cite{r2}}. These assumptions simplify the problem and cannot achieve the best performance. To tackle this problem, new optimization method should be applied for D2D communication underlaying HetNets.

Meanwhile, to enhance the security of wireless transmission, physical layer security has been developed based on information theoretic concepts\textcolor[rgb]{1.00,0.00,0.00}{\cite{n6,n7,n8}}. Physical-layer security, which exploits the physical characteristics of wireless channel to enhance the security, has significantly influenced the wireless communication because there exists a secured data transmission between power nodes and the eavsedropper\textcolor[rgb]{1.00,0.00,0.00}{\cite{7}}. To avoid the information from power nodes to UEs being leaked, the data rate of eavesdroppers should be maintained at a low level. With a more open control plane, D2D user equipment is more vulnerable to the attacks of eavesdroppers. There have been some studies to improve the secure capacity of D2D underlaying systems from the viewpoint of physical-layer security. In\textcolor[rgb]{1.00,0.00,0.00}{\cite{8}}, two secure capacity optimization problems for a multi-input and multi-output secrecy channel with multiple D2D communications are researched and two conservative approximation approaches to convert the probability based constraints into the deterministic constraint have been addressed. Furthermore, the radio resource allocation as a matching problem in a weighted bipartite graph has been formulated in \textcolor[rgb]{1.00,0.00,0.00}{\cite{9}}, in which the Kuhn-Munkres (KM) algorithm to obtain the optimal solution is proposed. However, the D2D communication only helps to improve the secure capacity of cellular UEs by confusing the eavesdroppers in both\textcolor[rgb]{1.00,0.00,0.00}{\cite{8}} and\textcolor[rgb]{1.00,0.00,0.00}{\cite{9}}. In other words, the D2D communication acts as a kind of interference against eavesdropping when an eavesdropper tries to overhear the cellular communication, therefore, the available rate of eavesdropper is lessened\textcolor[rgb]{1.00,0.00,0.00}{\cite{n5}}. To our best knowledge, there have been few publications to optimize the secure capacity of both cellular UEs and D2D links. In particular, few works maximize the secure capacity of the D2D communication underlaying HetNets. Note that such secure capacity optimization problem is non-convex and hard to be directly solved because of the existence of inter-tier and intra-tier interference.

The goal of this paper is to optimize the radio resource allocation aiming to maximize the secure capacity of D2D communication underlaying HetNets. The non-convex objective function with several constraints are first transformed into matrix form. Then, the equivalent convex form of the matrix form is derived according to the Perron-Frobenius theory which is based on the eigenvalue and eigenvector \textcolor[rgb]{1.00,0.00,0.00}{\cite{10,11,12}}. Note that the Perron-Frobenius theory has been mainly used in homogeneous networks, there are lack of publications to use this theory to address the optimization issue in the D2D communication underlaying HetNets. To derive this Perron-Frobenius theory based equivalent convex optimization problem, the corresponding Lagrange function has been formulated and the proximal operator has been evaluated. Finally, a novel iterative algorithm based on the proximal theory has been proposed. The main contributions of this paper can be summarized as following:

\begin{itemize}
 \item The optimization problem of radio resource allocation aiming to maximize the secure capacity in D2D communication underlaying HetNets has been addressed in this paper. To our best knowledge, previous works mainly take D2D communication as friendly jammer to improve the secure capacity of cellular users. Thus, only the performance of cellular users is optimized and the performance of D2D communication is ignored. Furthermore, previous works mainly focus on the traditional SE and do not focus on the secure capacity. Besides, when solving the optimization problem, previous works mainly try to simplify the interference between cellular users and D2D users by setting the interference with constant, which cannot take full advantage of the benefit of subcarrier reusing. This paper explores the resource allocation scheme for both cellular users and D2D users when they share the same subcarrier resource to maximize the secure capacity of D2D communication underlaying HetNets.
 \item To deal with the secrecy-optimized resource allocation problem, a non-convex objective function has been formulated. This non-convex objective function is hard to be directly derived. The existing works often simplify the interference into constant, and then the Lagrangian dual theory can be used. However, the interference in the D2D underlaying HetNets is often dynamical and should not be regarded as constant. To deal with this non-convex issue, the objective function is transformed to an equivalent convex optimization problem according to the Perron-Frobenius theory. Furthermore, to solve the equivalent convex problem, an iterative algorithm based on the proximal theory consisting of both outer and inner loop optimizations has been proposed to achieve the global optimal solution.
 \item The SE performance of the proposed resource allocation solution has been numerically evaluated. Simulation results show that the proposed iterative algorithm can promptly converge and outperforms the baseline algorithms. The effects of the QoS and power constraints with various other comparisons have been shown to evaluate the performance gains of the proposals.
\end{itemize}

The reminder of this paper is organized as follows. In Section II, the system model of D2D communication underlaying HetNets will be described and the optimization problem will be formulated. The solution to the optimization problem and the corresponding iterative algorithm will be presented in Section III. Section IV will provide simulations to verify
the effectiveness of the proposed algorithm and the corresponding solutions. Finally, the paper will be concluded in Section V.

\section{System Model}

\begin{figure}
\centering \vspace*{0pt}
\includegraphics[scale=0.38]{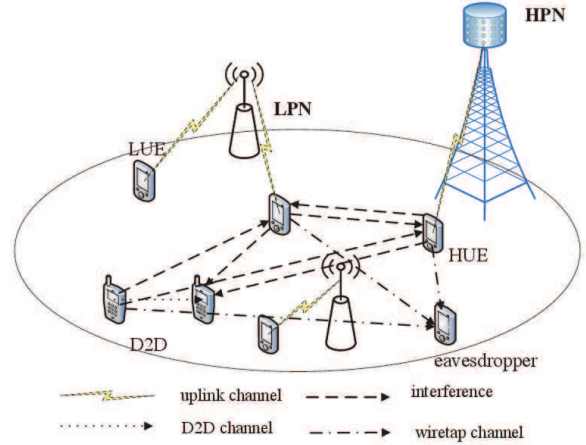}
\setlength{\belowcaptionskip}{-100pt} \caption{D2D communication underlaying HetNets.}
\label{System}\vspace*{-10pt}
\end{figure}

A D2D communication underlaying HetNets is shown in Fig. \ref{System}, in which $L$ LPNs sharing $N$ subcarriers with the HPN is considered. It is assumed that each subcarrier occupies $B$ MHz bandwidth. The HPN and LPNs are all closed accessed. The closed access means that each cell can only schedule the UEs that belong to it, while the open access means that one cell can schedule any UE as long as the reference signal receiving power is sufficiently high. Similar assumption can be found in previous works\textcolor[rgb]{1.00,0.00,0.00}{\cite{31,32}}. Under the closed access, UEs monopolize their own serving cells and the privacy and security of the communication can be guaranteed, though the performance of closed access is worse than that of open access. Since this paper focus on the secure capacity, the closed access is only considered. Let $H$ denote the average number of HPN-accessed UEs (HUEs) in each HPN and $M$ denote the average number of LPN-accessed UEs (LUEs) in each LPN, respectively. Besides, $K$ D2D-worked UEs (DUEs) co-exist with LUEs in each LPN. It is assumed that all UEs (e.g., HUEs, LUEs or DUEs) have low mobility so that the channel state information between UEs and power nodes as well as the channel state information between the transmitter and receiver of any D2D user pair remain stationary\cite{41,42,43}. Therefore, subcarriers can be assumed to be independent from each other, and the channel fading of each subcarrier can be assumed the same within a time slot, but may vary cross different subcarriers. On each subcarrier, there exists an eavesdropper that leverages the information of the desired UEs. Since DUEs prefer to share the uplink radio resources with HUEs or LUEs to avoid the severe inter-tier interference, this paper only focuses on the resource allocation to optimized the secure capacity in uplink.

\begin{table}[h]
\caption{Summary of Notations in This Paper}\label{centralized}
\scriptsize 
\centering
\begin{tabular}{p{0.5 in}p{2.6 in}}
 \hline
 HPN&high power node \\
 LPN&low power node \\
 HUE&HPN-accessed UE \\
 LUE&LPN-accessed UE \\
 DUE&D2D-worked UE\\

 ${h}_{hn}^{HH}$ &the channel gain from the $h$-th HUE to the HPN on the $n$-th subcarrier \\
 ${h}_{hnl}^{HL}$ &the channel gain from the $h$-th HUE to the $l$-th LPN on the $n$-th subcarrier \\
 ${h}_{hnlk}^{HD}$ &the channel gain from the $h$-th HUE to the receiver of the $k$-th DUE pair of the $l$-th LPN on the $n$-th subcarrier\\
 ${h}_{hn}^{HE}$ &the channel gain from the $h$-th HUE to the eavesdropper on the $n$-th subcarrier\\

 ${h}_{lmn}^{LH}$ &the channel gain from the $m$-th LUE of the $l$-th LPN to the HPN on the $n$-th subcarrier\\
 ${h}_{lmnj}^{LL}$ &the channel gain from the $m$-th LUE of the $l$-th LPN to the $j$-th LPN on the $n$-th subcarrier\\
 ${h}_{lmnjk}^{LD}$ &the channel gain from the $m$-th LUE of the $l$-th LPN to the receiver of the $k$-th DUE pair of the $j$-th LPN on the $n$-th subcarrier\\
 ${h}_{lmn}^{LE}$ &the channel gain from the $m$-th LUE of the $l$-th LPN to the eavesdropper on the $n$-th subcarrier\\

 ${h}_{lkn}^{DH}$ &the channel gain from the transmitter of the $k$-th DUE pair of the $l$-th LPN to the HPN on the $n$-th subcarrier\\
 ${h}_{lknj}^{DL}$ &the channel gain from the transmitter of the $k$-th DUE pair of the $l$-th LPN to the $j$-th LPN on the $n$-th subcarrier\\
 ${h}_{lmnjk}^{LD}$ &the channel gain from the transmitter of the $k$-th DUE pair of the $l$-th LPN to the receiver of the $i$-th DUE pair of the $j$-th LPN on the $n$-th subcarrier\\
 ${h}_{lmn}^{LE}$ &the channel gain from the transmitter of the $k$-th DUE pair of the $l$-th LPN to to the eavesdropper on the $n$-th subcarrier\\

 $p_{hn}^H$ &the transmit power of the $h$-th HUE on the $n$-th subcarrier\\
 $p_{lmn}^L$ &the transmit power of the $m$-th LUE of the $l$-th LPN on the $n$-th subcarrier\\
 $p_{lkn}^D$ &the transmitter of the $k$-th DUE pair of the $l$-th LPN on the $n$-th subcarrier\\

 $\alpha _{hn}^H$&subcarrier allocation index which denotes whether the $n$-th subcarrier is allocated to the $h$-th HUE\\
 $\alpha _{lmn}^L$&subcarrier allocation index which denotes whether the $n$-th subcarrier is allocated to the $m$-th LUE of the $l$-th LPN\\
 $\alpha _{lkn}^D$ &subcarrier allocation index which denotes whether the $n$-th subcarrier is allocated to the $k$-th DUE pair of the $l$-th LPN\\

 \hline
\end{tabular}
\end{table}

Let ${h}_{hn}^{HH}, {h}_{hnl}^{HL}, {h}_{hnlk}^{HD}$, and ${h}_{hn}^{HE}$ denote the radio channel gain from the $h$-th HUE to the HPN, from the $h$-th HUE to the $l$-th LPN, from the $h$-th HUE to the receiver of the $k$-th DUE pair of the $l$-th LPN, and from the $h$-th HUE to the eavesdropper on the $n$-th subcarrier, respectively. Meanwhile, ${h}_{lmn}^{LH}$, ${h}_{lmnj}^{LL}$, ${h}_{lmnjk}^{LD}$, and ${h}_{lmn}^{LE}$ denote the radio channel gain from the $m$-th LUE of the $l$-th LPN to the HPN, from the $m$-th LUE of the $l$-th LPN to the $j$-th LPN, from the $m$-th LUE of the $l$-th LPN to the receiver of the $k$-th DUE pair of the $j$-th LPN, and from the $m$-th LUE of the $l$-th LPN to the eavesdropper on the $n$-th subcarrier, respectively. Similarly, let ${h}_{lkn}^{DH}$, ${h}_{lknj}^{DL}$, ${h}_{lknji}^{DD}$, and ${h}_{lkn}^{DE}$ denote the channel gain from the transmitter of the $k$-th DUE pair of the $l$-th LPN to the HPN, to the $j$-th LPN, to the receiver of the $i$-th DUE pair of the $j$-th LPN, and to the eavesdropper on the $n$-th subcarrier, respectively. Denote $p_{hn}^H$, $p_{lmn}^L$, and $p_{lkn}^D$ by the transmit power of the $h$-th HUE, the transmit power of the $m$-th LUE of the $l$-th LPN, and the transmitter of the $k$-th DUE pair of the $l$-th LPN on the $n$-th subcarrier, respectively. In orthogonal frequency division multiple access (OFDMA) based wireless networks, a base station allocates orthogonal subcarriers to different users. Thus, in each cell (e.g., an HPN or a LPN), each subcarrier can be allocated to only one cellular UE or a DUE pair\cite{111,222,333}, but one or more subcarriers can be allocated to the same cellular UE or a DUE pair. It means that at most one element of $\left( {p_{h1}^H,p_{h2}^H, \cdots p_{hN}^H} \right)$ is positive and the others are all zero. Let $\alpha _{hn}^H,\alpha _{lmn}^L,\alpha _{lkn}^D \in \{ 0,1\}$ denote the allocation index of subcarriers, which denotes whether the $n$-th subcarrier is allocated to the $h$-th HUE, the $m$-th LUE of the $l$-th LPN, and the $k$-th DUE pair of the $l$-th LPN or not, respectively. It should be noted that eavesdroppers are passive and will not share its CSI with the HPN or LPN. In fact, if eavesdroppers keep silent all the time, it is impossible for the network to estimate the CSI of them. However, when eavesdropper become active, they can be detected by the network\textcolor[rgb]{1.00,0.00,0.00}{\cite{detected}}. Thus, in this paper, the CSI of eavesdropper is assumed to be known and similar assumption can be found in previous works\textcolor[rgb]{1.00,0.00,0.00}{\cite{fullCSI1,fullCSI2,fullCSI3}}.

The uplink signal-to-interference-plus-noise ratio (SINR)from the $h$-th HUE to the HPN on the $n$-th subcarrier $\rho _{hn}^H$ can be expressed as:

\begin{equation}\label{eqn_dbl_x}
\rho _{hn}^H = \frac{{\alpha _{hn}^Hp_{hn}^H{\rm{h}}_{hn}^{HH}}}{{\sum\limits_{l = 1}^L {\sum\limits_{m = 1}^M {\alpha _{lmn}^Lp_{lmn}^L{\rm{h}}_{lmn}^{LH}} } + \sum\limits_{l = 1}^L {\sum\limits_{k = 1}^K {\alpha _{lkn}^Dp_{lkn}^D{\rm{h}}_{lkn}^{DH}} } + B{\sigma ^2}}}.
\end{equation}

The uplink SINR from the $h$-th HUE to the eavesdropper on the $n$-th subcarrier $\rho _{hn}^{HE}$ can be expressed as:

\begin{equation}
\rho _{hn}^{HE} = \frac{{\alpha _{hn}^Hp_{hn}^H{\rm{h}}_{hn}^{HE}}}{{\sum\limits_{l = 1}^L {\sum\limits_{m = 1}^M {\alpha _{lmn}^Lp_{lmn}^L{\rm{h}}_{ln}^{LE}} } + \sum\limits_{l = 1}^L {\sum\limits_{k = 1}^K {\alpha _{lkn}^Dp_{lkn}^D{\rm{h}}_{lkn}^{DE}} } + B{\sigma ^2}}}.
\end{equation}

The uplink SINR from the $m$-th LUE of the $l$-th LPN to the $l$-th LPN on the $n$-th subcarrier $\rho _{lmn}^L$ can be expressed as:
\begin{equation}
\rho _{lmn}^L = \frac{{\alpha _{lmn}^Lp_{lmn}^L{\rm{h}}_{lmnl}^{LL}}}{{I_{lmn}^L + {\sigma ^2}}},
\end{equation}
where $I_{lmn}^L$ denotes the interference and is expressed as:
\begin{eqnarray}
I_{lmn}^L{\rm{ = }}&&\sum\limits_{h = 1}^H {\alpha _{hn}^Hp_{hn}^H{\rm{h}}_{hnl}^{HL}} + \sum\limits_{j = 1,j \ne l}^L {\sum\limits_{i = 1}^M {\alpha _{jin}^Lp_{jin}^L{\rm{h}}_{jinl}^{LL}} } \nonumber\\
&&+ \sum\limits_{j = 1}^L {\sum\limits_{k = 1}^K {\alpha _{jkn}^Dp_{jkn}^D{\rm{h}}_{jknl}^{DL}} } .
\end{eqnarray}

The uplink SINR from the the $l$-th LPN to the eavesdropper on the $n$-th subcarrier $\rho _{lmn}^{LE}$ can be expressed as:
\begin{equation}\label{eqn_dbl_y}
\rho _{lmn}^{LE}{\rm{ = }}\frac{{\alpha _{lmn}^Lp_{lmn}^L{\rm{h}}_{lmn}^{LE}}}{{I_{lmn}^{LE} + {\sigma ^2}}},
\end{equation}
where $I_{lmn}^{LE}$ denotes the interference and is expressed as:
\begin{eqnarray}
I_{lmn}^{LE} =&& \sum\limits_{h = 1}^H {\alpha _{hn}^Hp_{hn}^H{\rm{h}}_{hn}^{HE}} + \sum\limits_{j = 1,j \ne l}^L {\sum\limits_{i = 1}^M {\alpha _{jin}^Lp_{jin}^L{\rm{h}}_{jn}^{LE}} } \nonumber\\
&& + \sum\limits_{j = 1}^L {\sum\limits_{k = 1}^K {\alpha _{jkn}^Dp_{jkn}^D{\rm{h}}_{jkn}^{DE}} } .
\end{eqnarray}

The uplink SINR from the transmitter of the $k$-th DUE pair of the $l$-the LPN to the receiver of the $k$-th DUE pair of the $l$-the LPN on the $n$-th subcarrier $\rho _{lkn}^D$ can be expressed as:
\begin{equation}
\rho _{lkn}^D{\rm{ = }}\frac{{\alpha _{lkn}^Dp_{lkn}^D{\rm{h}}_{lknlk}^{DD}}}{{I_{lkn}^D + {\sigma ^2}}},
\end{equation}
where $I_{lkn}^D$ denotes the interference and is expressed as:
\begin{eqnarray}
I_{lkn}^D = &&\sum\limits_{h = 1}^H {\alpha _{hn}^Hp_{hn}^H{\rm{h}}_{hnl}^{HD}} + \sum\limits_{j = 1}^L {\sum\limits_{m = 1}^M {\alpha _{jmn}^Lp_{jmn}^L{\rm{h}}_{jmnl}^{LD}} } \nonumber\\
&& + \sum\limits_{j = 1,j \ne l}^L {\sum\limits_{i = 1}^K {\alpha _{jin}^Dp_{jin}^D{\rm{h}}_{jinlk}^{DD}} } .
\end{eqnarray}

Similarly, the uplink SINR from the transmitter of the $k$-th DUE pair of the $l$-the LPN to the eavesdropper $\rho _{lkn}^{DE}$ can be expressed as:
\begin{equation}\label{eqn_dbl_a}
\rho _{lkn}^{DE}{\rm{ = }}\frac{{\alpha _{lkn}^Dp_{lkn}^D{\rm{h}}_{lkn}^{DE}}}{{I_{lkn}^{DE} + {\sigma ^2}}},
\end{equation}
where $I_{lkn}^{DE}$ denotes the interference and is expressed as:
\begin{eqnarray}
I_{lkn}^{DE}{\rm{ = }}&&\sum\limits_{h = 1}^H {\alpha _{hn}^Hp_{hn}^H{\rm{h}}_{hn}^{HE}} + \sum\limits_{j = 1}^L {\sum\limits_{m = 1}^M {\alpha _{jmn}^Lp_{jmn}^L{\rm{h}}_{ln}^{LE}} }\nonumber\\
&& + \sum\limits_{j = 1,j \ne l}^L {\sum\limits_{i = 1}^K {\alpha _{jin}^Lp_{jin}^D{\rm{h}}_{jin}^{DE}} } .
\end{eqnarray}

Let $C_{hn}^{HS}$, $C_{lmn}^{LS}$, $C_{lkn}^{DS}$ denote the secrecy capacity of the $h$-th HUE, the $m$-th LUE of the $l$-th LPN, and the $k$-th DUE pair of the $l$-the LPN, respectively. Accordingly, $C_{hn}^{HS}$, $C_{lmn}^{LS}$, $C_{lkn}^{DS}$ can be expressed as:
\begin{eqnarray}\label{eqn:rates}
\label{eqn_dbl_x}
&&C_{hn}^{HS} = B\log (1 + \rho _{hn}^H) - B\log (1 + \rho _{hn}^{HE}),\\
&&C_{lmn}^{LS}= B\log (1 + \rho _{lmn}^L)- B\log (1 + \rho _{lmn}^{LE}),\\
&&C_{lkn}^{DS}= B\log (1 + \rho _{lkn}^D)- B\log (1 + \rho _{lkn}^{DE}),
\end{eqnarray}
respectively. Therefore, the secrecy capacity of the D2D communication underlaying HetNets can be written as:
\begin{equation}
\sum\limits_{h = 1}^H {\sum\limits_{n = 1}^N {C_{hn}^{HS}} } + \sum\limits_{l = 1}^L {\sum\limits_{m = 1}^M {\sum\limits_n^N {C_{lmn}^{LS}} } } + \sum\limits_{l = 1}^L {\sum\limits_{k = 1}^K {\sum\limits_{n = 1}^N {C_{lkn}^{DS}} } }.
\end{equation}

It is assumed that the secrecy capacity of each UE should be no less than the pre-defined threshold to guarantee QoS. Meanwhile, to save the energy consumption, the maximum allowed transmit power of each power node is limited. Besides, the subcarrier allocation index should be restricted to prevent the same subcarrier being allocated to more than one UE of the same type, e.g., HUEs, LUEs, and DUEs. As a result, the secrecy-optimized resource allocation problem for the D2D communication underlaying HetNets can be formulated with the QoS and transmit power constraints.

\begin{prob}[\textbf{Secrecy Capacity Optimization}] With the
constraints on the required QoS and maximum
transmit power allowance, the secrecy-optimized resource allocation problem for the D2D communication underlaying HetNets can be formulated as:
\begin{eqnarray}
\mathop {{\rm{max}}}\limits_{\{ p_{hn}^H,p_{lmn}^L,p_{lkn}^D\} }
&&\sum\limits_{h = 1}^H {\sum\limits_{n = 1}^N {C_{hn}^{HS}} } + \sum\limits_{l = 1}^L {\sum\limits_{m = 1}^M {\sum\limits_n^N {C_{lmn}^{LS}} } } \nonumber\\
&&+ \sum\limits_{l = 1}^L {\sum\limits_{k = 1}^K {\sum\limits_{n = 1}^N {C_{lkn}^{DS}} } } \label{OP1}\\
{\rm{s.t.}}&&\sum\limits_{n = 1}^N {C_{hn}^{HS}} \ge C_{\min }^{HS}{\rm{ }}, \forall h,\label{cons1}\\
&&\sum\limits_n^N {C_{lmn}^{LS}} \ge C_{\min }^{LS}{\rm{ }}, \forall l,m,\label{cons2}\\
&&\sum\limits_{n = 1}^N {C_{lkn}^{DS}} \ge C_{\min }^{DS}{\rm{ }}, \forall l,k,\label{cons3}\\
&&p_{hn}^H \le P_{\max }^H,\forall h,n\label{cons4}\\
&&p_{lmn}^L \le P_{\max }^L,\forall l,m,n\label{cons5}\\
&&p_{lkn}^D \le P_{\max }^D,\forall l,k,n\label{cons6}\\
&&\sum\limits_{h = 1}^H {\alpha _{hn}^H} = 1{\rm{ }}, \forall n,{\rm{ }}\alpha _{hn}^H \in \{ 0,1\} ,\label{cons7}\\
&&\sum\limits_{m = 1}^M {\alpha _{lmn}^L} = 1{\rm{ }}, \forall l,n,{\rm{ }}\alpha _{lmn}^L \in \{ 0,1\} ,\label{cons8}\\
&&\sum\limits_{k = 1}^K {\alpha _{lkn}^D} = 1{\rm{ }}, \forall l,n,{\rm{ }}\alpha _{lkn}^D \in \{ 0,1\} ,\label{cons9}
\end{eqnarray}
where the constraints \eqref{cons1}-\eqref{cons3} correspond to the required lower bound of the the secrecy capacity of HUEs, LUEs, and DUEs, respectively. The constraints \eqref{cons4}-\eqref{cons6} represent the maximum allowed transmit power of HUEs, LUEs, and DUEs, respectively. The constraints \eqref{cons7}-\eqref{cons9} put limitations on the subcarrier allocation policy so that each subcarrier can only be allocated to at most one UE of the same type.
\end{prob}

\section{Secrecy-Optimized Resource Allocation Optimization}

Joint optimization of subcarriers and transmit power is an intractable problem\textcolor[rgb]{1.00,0.00,0.00}{\cite{5}}, and the optimization problem \eqref{OP1} with constraints \eqref{cons1}-\eqref{cons9} listed in Section II is non-deterministic polynomial (NP) hard, which is impossible to be directly solved. But it is noted that under any given the subcarrier allocation policy, problem (15) can be transformed to a convex form according to the Perron-Frobenius theory. Thus, if we solve the equivalent convex problem for every possible subcarrier allocation policy, the global optimal solution can be derived. However, such exhaustive method with high complexity is hard to be practical. In this paper, we solve this transformed convex-optimization problem with two steps. In the first step, we propose a suboptimal algorithm to allocate each user with proper subcarriers. In the second step, we design an optimal power allocation scheme under the subcarrier allocation policy proposed in the first step. Similar method to tackle the NP-hard problem has been used in previous works\textcolor[rgb]{1.00,0.00,0.00}{\cite{NP1,NP2,NP3}}. Based on the proposed subcarrier allocation solution, it can be demonstrated that the optimization problem \eqref{OP1} can be completely solved through proposing the optimal power allocation solution.

\subsection{Subcarrier Allocation}
In this subsection, subcarrier allocation scheme is derived for solving problem \eqref{OP1}. Note that the subcarrier allocation index is a discrete variable and for any given subcarrier allocation scheme the optimal power allocation is independent from the subcarrier allocation results. Therefore, the following theorem can be holden.

\textbf{Theorem 1:} For any optimization problem $ \mathop {{\rm{max}}}\limits_{\{ \bf{x},\bf{y}\} } f(\bf{x},\bf{y})$, where ${\bf{x}} \in {\bf{N}^{1 \times X}}$ is a finite discrete variable vector and each element of $\bf{x}$ is one of $S$ integer values. ${\bf{y}} \in {{\bf{R}}^{1 \times Y}}$ is continuous variable. $X$, $S$ and $Y$ are positive integers. Let ${{{\bf{x}}_i}}$ denote the $i$-th feasible solution of $\bf{x}$, where $\bf{x}^*$ is the optimal $\bf{x}$, and ${\bf{y}}_i^*$ denote the optimal ${\bf{y}}$ under ${{{\bf{x}}_i}}$ and $1 \le i \le {X^{S}}$. Therefore, $\bf{x}^*$ can be expressed as:
\begin{eqnarray}
{{\bf{x}}^*} = \mathop {{\rm{argmax}}}\limits_{\{ {{\bf{x}}_i}\} } {\kern 1pt} {\kern 1pt} {\kern 1pt} {\kern 1pt} {\kern 1pt} \left( {f({{\bf{x}}_i},{\bf{y}}_i^*){\kern 1pt} } \right).
\end{eqnarray}
\hfill\rule{8pt}{8pt}

\textbf{\emph{Proof:}} Please see Appendix A. \hfill\rule{8pt}{8pt}

According to \emph{Theorem 1}, let $\bf{x}$ denote the subcarrier allocation scheme and $\bf{y}$ denote the transmit power allocation scheme, the optimal subcarrier allocation scheme can be directly derived with exhaustive method. However, such exhaustive method is impractical because each subcarrier needs to be searched to derive the optimal subcarrier allocation, which resulting in a high complexity. In this subsection, a suboptimal subcarrier allocation scheme is proposed to optimize the capacity of each cell (e.g., HPNs or LPNs), which can be expressed as:
\begin{eqnarray}
\mathop {{\rm{max}}}
&&\sum\limits_{h = 1}^H {\sum\limits_{n = 1}^N {\log (1 + \frac{{\alpha _{hn}^Hp_{hn}^H{\rm{h}}_{hn}^{HH}}}{{{B\sigma ^2}}})} } \\
{\rm{s.t.}}&&\sum\limits_{h = 1}^H {\sum\limits_n^N {p_{hn}^H{\rm{h}}_{hnl}^{HL}} } \le {I^H},\\
&&\sum\limits_{h = 1}^H {\alpha _{hn}^H} = 1{\rm{ }}\forall n,{\rm{ }}\alpha _{hn}^H \in \{ 0,1\} ,\\
{\rm{var}}&&p_{hn}^H,\alpha_{hn}^H,
\end{eqnarray}
where ${I^H}$ is the maximum allowed interference caused by HUEs.

Obviously, it can be solved by using the Lagrangian function, i.e.,
\begin{align}
L(\alpha _{hn}^H,p_{hn}^H) = \sum\limits_{h = 1}^H {\sum\limits_{n = 1}^N {\log (1 + \frac{{\alpha _{hn}^Hp_{hn}^H{\rm{h}}_{hn}^{HH}}}{{{\sigma ^2}}})} }\nonumber\\- \lambda (\sum\limits_{h = 1}^H {\sum\limits_n^N {p_{hn}^H{\rm{h}}_{hnl}^{HL}} } - {I^H}),
\end{align}
where $\lambda$ is the Lagrange multiplier.

According to the Karush-Kuhn-Tucker (KKT) condition, the optimal subcarrier allocation scheme to maximize the data rate of HPNs can be expressed as:
\begin{equation}\label{srate:HUE}
\alpha _{hn}^H = \left\{ {\begin{array}{*{20}{c}}
1&{{\rm{n = arg}}\mathop {{\rm{max}}}\limits_n {\rm{ log(}}\frac{1}{{\lambda {\sigma ^2}}}{\rm{) - }}\lambda {{\left[ {\frac{1}{\lambda } - {\sigma ^2}} \right]}^ + }}\\
0&{{\rm{others}}}
\end{array}} \right..
\end{equation}

Initialize the Lagrange multiplier and then use the sub-gradient-based
method\textcolor[rgb]{1.00,0.00,0.00}{\cite{b8}} to search for the subcarrier allocation scheme. The proposed algorithm is executed in each cell (e.g., HPNs or LPNs) to a achieve good performance. In the whole network comprised of multiple HPNs, this proposal can work efficiently in each HPN, thus the overall performance can be optimized.

\subsection{Power Allocation Scheme}

In this subsection, based on the aforementioned subcarrier allocation, the power allocation to solve problem \eqref{OP1} is researched. Problem \eqref{OP1} can be rewritten as:

\begin{eqnarray}
\sum\limits_{h = 1}^H {\sum\limits_{n = 1}^N {C_{hn}^{HS}} } + \sum\limits_{l = 1}^L {\sum\limits_{m = 1}^M {\sum\limits_n^N {C_{lmn}^{LS}} } } + \sum\limits_{l = 1}^L {\sum\limits_{k = 1}^K {\sum\limits_{n = 1}^N {C_{lkn}^{DS}} } }=\nonumber\\
\sum\limits_{n = 1}^N {\left( {\sum\limits_{h = 1}^H {C_{hn}^{HS}} + \sum\limits_{l = 1}^L {\sum\limits_{m = 1}^M {C_{lmn}^{LS}} } + \sum\limits_{l = 1}^L {\sum\limits_{k = 1}^K {C_{lkn}^{DS}} } } \right)}.
\end{eqnarray}

In OFDMA based wireless networks, the interference between any two different subcarriers can be ignored due to the orthogonal characteristics. Therefore, the primal objective function can be divided into $N$ independent subproblems, which indicates that the primal problem can be derived by solving the $N$ subproblems. The $n$-th subproblem optimization can be given as:
\begin{eqnarray}
\mathop {{\rm{max}}}\limits_{\{ p_{hn}^H,p_{lmn}^L,p_{lkn}^D\} }
&&\sum\limits_{h = 1}^H {C_{hn}^{HS}} + \sum\limits_{l = 1}^L {\sum\limits_{m = 1}^M {C_{lmn}^{LS}} } + \sum\limits_{l = 1}^L {\sum\limits_{k = 1}^K {C_{lkn}^{DS}} }\nonumber\\\label{eqn:sub1} \\
{\rm{s.t.}}&&C_{hn}^{HS} \ge C_{\min }^{HS}{\rm{ }}, \forall h, \\
&&C_{lmn}^{LS} \ge C_{\min }^{LS}{\rm{ }}, \forall l,m, \\
&&C_{lkn}^{DS} \ge C_{\min }^{DS}{\rm{ }}, \forall l,k, \\
&&p_{hn}^H \le P_{\max }^H,\forall h,n, \\
&&p_{lmn}^L \le P_{\max }^L,\forall l,m,n, \\
&&p_{lkn}^D \le P_{\max }^D,\forall l,k,n.
\end{eqnarray}

Note that \eqref{eqn:sub1} is still non-convex due to the inter-tier interference. To transform it into a convex form, the uplink channel state information (CSI) vector of the $h$-th HUE on the $n$-th subcarrier, which represents the CSI from the $h$-th HUE to the HPN, from the $h$-th HUE to LPNs, and from the $h$-th HUE to the receiver of other DUEs that share the $n$-th subcarrier, is given by
\begin{equation}\label{g1}
{\bf{G}}_{hn}^H=\left[ {{\rm{h}}_{hn}^{HH}{\rm{,h}}_{hn1}^{HL}, \cdots {\rm{h}}_{hnl}^{HL} \cdots ,{\rm{h}}_{hn1}^{HD}, \cdots {\rm{h}}_{hnl}^{HD} \cdots } \right]_{}^{\rm T},
\end{equation}
where $h,m,l \in \left\{ {h,l|\exists h,m,l{\rm{ }}\alpha _{hn}^H,\alpha _{lmn}^L,\alpha _{lkn}^D \ne 0} \right\}$. Accurately, under given subcarrier allocation scheme, at most one LUE can be allocated to the $n$-th subcarrier of the $l$-th LPN. Similarly, the uplink CSI vector for the LUE on the $n$-th subcarrier of the $l$-th LPN can be given as:
\begin{equation}
{\bf{G}}_l^L = \left[ {{\rm{h}}_{lmn}^{LH}{\rm{,h}}_{lmn1}^{LL}, \cdots {\rm{h}}_{lmnl}^{LL} \cdots ,{\rm{h}}_{lmn1}^{LD}, \cdots {\rm{h}}_{lmnl}^{LD} \cdots } \right]_{}^{\rm T},
\end{equation}
where $h,m,l \in \left\{ {h,l|\exists h,m,l{\rm{ }}\alpha _{hn}^H,\alpha _{lmn}^L,\alpha _{lkn}^D \ne 0} \right\}$. Similarly, the uplink CSI vector for the DUE on the $n$-th subcarrier of the $l$-th LPN can be given as:
\begin{equation}
{\bf{G}}_l^D = \left[ {{\rm{h}}_{lkn}^{DH}{\rm{,h}}_{lkn1}^{DL}, \cdots {\rm{h}}_{lknl}^{DL} \cdots ,{\rm{h}}_{lkn1}^{DD} \cdots ,{\rm{h}}_{lknl}^{DD} \cdots } \right]_{}^{\rm T},
\end{equation}
where $h,k,l \in \left\{ {h,l|\exists h,k,l{\rm{ }}\alpha _{hn}^H,\alpha _{lmn}^L,\alpha _{lkn}^D \ne 0} \right\}$.

Combine these three CSIs ${\bf{G}}_{hn}^H$, ${\bf{G}}_l^L$, ${\bf{G}}_l^D$ together, it can be derived that on the $n$-th subcarrier, the CSI matrix among all users on the $n$-th subcarrier can be expressed as:
 \begin{equation}
{{\bf{G}}_n} = \left[ {{\bf{G}}_{hn}^H,\underbrace {{\bf{G}}_1^L,{\bf{G}}_2^L, \cdots {\bf{G}}_l^L, \cdots }_{{\rm{LPN}}}\underbrace {{\bf{G}}_1^D,{\bf{G}}_2^D \cdots {\bf{G}}_l^D \cdots }_{{\rm{D2D }}user pair}{\bf{G}}_n^E} \right].
\end{equation}

Similarly, the CSI between users and the eavesdropper on the $n$-th subcarrier is given by
\begin{align}
&{\bf{G}}_n^E = \nonumber\\
&\left[ {{\rm{h}}_{hn}^{HE},\underbrace {{\rm{h}}_{1mn}^{LE},{\rm{h}}_{2mn}^{LE}, \cdots {\rm{h}}_{lmn}^{LE}, \cdots }_{LPN}\underbrace {{\rm{h}}_{1kn}^{DE},{\rm{h}}_{2kn}^{DE}, \cdots {\rm{h}}_{lkn}^{DE} \cdots }_{D2D}} \right].
\end{align}

Note that for a given subcarrier allocation scheme, ${{\bf{G}}_n}$ and ${\bf{G}}_n^E$ are fixed, which means that the number of users (HUE, LUE and DUE) on the $n$-th subcarrier is determined and the CSI matrix can be known. Without loss of generality, let ${{\bf{G}}_n}$ be a $J \times J$ matrix which denotes that the number of users on the $n$-th channel is $J$, and let ${\bf{G}}_n^E$ be a $J\times 1$ vector. The transmit power matrix of the $J$ users is given by ${{\bf{p}}_n}$. A nonnegative matrix is defined ${{\bf{F}}_n}$ with entries
\begin{equation}
{\left( {{{\bf{F}}_n}} \right)_{ij}} = \left\{ {\begin{array}{*{20}{c}}
0&{{\rm{if }} i = j{\rm{ }}}\\
{\frac{{{{\left( {{{\bf{G}}_n}} \right)}_{ij}}}}{{{{\left( {{{\bf{G}}_n}} \right)}_{jj}}}}}&{{\rm{if }} i \ne j}
\end{array}} \right.,
\end{equation}
and the vector ${{\bf{v}}_n} = {\left( {\frac{{{\sigma ^2}}}{{{{\left( {{{\bf{G}}_n}} \right)}_{11}}}},\frac{{{\sigma ^2}}}{{{{\left( {{{\bf{G}}_n}} \right)}_{22}}}}, \cdots \frac{{{\sigma ^2}}}{{{{\left( {{{\bf{G}}_n}} \right)}_{JJ}}}}} \right)^{\rm T}}$.

Now, the SINR of the $j$-th user in the uplink is ${\left( {{{\bf{S}}_n}} \right)_j} = \frac{{{p_{nj}}}}{{{{\left( {{{\bf{F}}_n}{{\bf{p}}_n} + {{\bf{v}}_n}} \right)}_j}}}$, and its corresponding capacity can be expressed as:
\begin{equation}\label{append1}
{\left( {{{\bf{C}}_n}} \right)_j} = \log \left( {1 + {{\left( {{{\bf{S}}_n}} \right)}_j}} \right).
\end{equation}

Similarly, we can have
\begin{equation}
{\left( {{\bf{F}}_n^E} \right)_{ij}} = \left\{ {\begin{array}{*{20}{c}}
0&{{\rm{if }}i = j}\\
{\frac{{{{\left( {{\bf{G}}_n^E} \right)}_i}}}{{{{\left( {{\bf{G}}_n^E} \right)}_j}}}}&{{\rm{if }}i \ne j}
\end{array}} \right.,
\end{equation}
and the vector ${\bf{v}}_n^E = {\left( {\frac{{{\sigma ^2}}}{{{{\left( {{\bf{G}}_n^E} \right)}_1}}},\frac{{{\sigma ^2}}}{{{{\left( {{\bf{G}}_n^E} \right)}_2}}}, \cdots \frac{{{\sigma ^2}}}{{{{\left( {{\bf{G}}_n^E} \right)}_J}}}} \right)^{\rm T}}$.

Now the SINR of the $j$-th user in the wiretap channel can be expressed as ${\left( {{\bf{S}}_n^E} \right)_j} = \frac{{{p_{nj}}}}{{{{\left( {{\bf{F}}_n^E{{\bf{p}}_n} + {\bf{v}}_n^E} \right)}_j}}}$, and the data rate is ${\left( {{\bf{C}}_n^E} \right)_j} = \log \left( {1 + {{\left( {{\bf{S}}_n^E} \right)}_j}} \right)$. Rewrite \eqref{eqn:sub1} as:
\begin{eqnarray}
\mathop {{\rm{max}}}\limits_{\{ {{\bf{p}}_n}\} }
&&\sum\limits_{j = 1}^J {{{\left( {{{\bf{C}}_n}} \right)}_j}} - \sum\limits_{j = 1}^J {{{\left( {{\bf{C}}_n^E} \right)}_j}} \label{eqn:sub2} \\
{\rm{s.t.}}&&{\left( {{{\bf{C}}_n}} \right)_j} - {\left( {{\bf{C}}_n^E} \right)_j} \ge {\left( {{\bf{C}}_{\min }^S} \right)_j}{\rm{ }}, \forall j, \\
&&0 \le {\left( {{{\bf{p}}_n}} \right)_j} \le {\left( {{{\bf{p}}_{\max }}} \right)_j}{\rm{ }}, \forall j,\label{eqn:cons1}
\end{eqnarray}
where ${\bf{C}}_{\min }^S$ and ${{\bf{p}}_{\max }}$ are the minimum secrecy data rate and maximum transmit power constraints, respectively. Due to the inter-tier interference, the optimization problem in \eqref{eqn:sub2} is still hard to be derived. To solve this problem, we can define the vector ${\bm{\alpha }_j}$
\begin{equation}
{\left( {{{{\bm{\alpha }}}_j}} \right)_i} = \left\{ {\begin{array}{*{20}{c}}
0&{i \ne j}\\
1&{i = j}
\end{array}} \right.
\end{equation}
and the constraint \eqref{eqn:cons1} can be expressed as:
\begin{equation}\label{a}
{\bm{\alpha }}_j^{\rm T}{{\bf{p}}_n} \le {\left( {{{\bf{p}}_{\max }}} \right)_j}{\rm{ }}\forall j{\rm{ }}.
\end{equation}

To transfer the constraint \eqref{eqn:cons1}, we can have the following theorem:

\textbf{Theorem 2:} \emph{Let}

\begin{eqnarray}
&&{{\bf{B}}_{nj}} = { {{{\bf{F}}_n}} } + \frac{1}{{{{\left( {{{\bf{p}}_{\max }}} \right)}_j}}}{{\bf{v}}_n}{\bm{\alpha }}_j^{\rm T},\label{b} \\
&&{\bf{B}}_{nj}^E = { {{\bf{F}}_n^E} } + \frac{1}{{{{\left( {{{\bf{p}}_{\max }}} \right)}_j}}}{\bf{v}}_n^E{\bm{\alpha }}_j^{\rm T}, \label{be}\\
&&{{\bf{q}}_n} = {{\bf{F}}_n}{{\bf{p}}_n} + {{\bf{v}}_n},\label{q} \\
&&{\bf{q}}_n^E = {\bf{F}}_n^E{{\bf{p}}_n} + {\bf{v}}_n^E,\label{qe}
\end{eqnarray}
\emph{then constraint \eqref{eqn:cons1} is equal to:}
\begin{eqnarray}
&&{{\bf{B}}_{nj}}{\rm{diag}}({e^{{{\bf{C}}_n}}}){{\bf{q}}_n} \le \left( {{\bf{I}} + {{\bf{B}}_{nj}}} \right){{\bf{q}}_n}{\rm{ }}\forall j,\label{bcons1} \\
&&{\bf{B}}_{nj}^E{\rm{diag}}({e^{{\bf{C}}_n^E}}){\bf{q}}_n^E \le \left( {{\bf{I}} + {\bf{B}}_{nj}^E} \right){\bf{q}}_n^E{\rm{ }}\forall j.\label{bcons2}
\end{eqnarray}
\hfill\rule{8pt}{8pt}

\textbf{\emph{Proof:}} Please see Appendix B. \hfill\rule{8pt}{8pt}

Based on \emph{Theorem 2}, the optimization problem in \eqref{eqn:sub2} can be transformed to
\begin{eqnarray}
\mathop {{\rm{max}}}\limits_{\{ {{\bf{C}}_n},{\bf{C}}_n^E,{{\bf{q}}_n},{\bf{q}}_n^E\} }
&&\sum\limits_{j = 1}^J {{{\left( {{{\bf{C}}_n}} \right)}_j}} - \sum\limits_{j = 1}^J {{{\left( {{\bf{C}}_n^E} \right)}_j}} \label{eqn:sub3} \\
{\rm{s.t.}}&&{\left( {{{\bf{C}}_n}} \right)_j} - {\left( {{\bf{C}}_n^E} \right)_j} \ge {\left( {{\bf{C}}_{\min }^S} \right)_j}{\rm{ }}, \forall j, \\
&&{{\bf{B}}_{nj}}{\rm{diag}}({e^{{{\bf{C}}_n}}}){{\bf{q}}_n} \le \left( {{\bf{I}} + {{\bf{B}}_{nj}}} \right){{\bf{q}}_n}{\rm{ }}, \forall j,\nonumber\\\label{eqn:cons2}\\
&&{\bf{B}}_{nj}^E{\rm{diag}}({e^{{\bf{C}}_n^E}}){\bf{q}}_n^E \le \left( {{\bf{I}} + {\bf{B}}_{nj}^E} \right){\bf{q}}_n^E{\rm{ }}, \forall j.\nonumber\\\label{eqn:cons3}
\end{eqnarray}

Obviously, the objective function \eqref{eqn:sub3} has been transferred to be linear. However, this optimization problem is still non-convex because of the non-convex constraints \eqref{eqn:cons2} and \eqref{eqn:cons3}. These two constraints can be transformed into be convex by using the following theorem.

\textbf{Theorem 3:} \emph{If the non-negative matrix ${{{\bf{\tilde B}}}_{nj}} = {\left( {{\bf{I}} + {{\bf{B}}_{nj}}} \right)^{ - 1}}{{\bf{B}}_{nj}}$ and ${\bf{\tilde B}}_{nj}^E = {\left( {{\bf{I}} + {\bf{B}}_{nj}^E} \right)^{ - 1}}{\bf{B}}_{nj}^E$ holds, the optimization problem in \eqref{eqn:sub3} can be transformed into a convex form as:}
\begin{eqnarray}
\mathop {{\rm{max}}}\limits_{\{ {{\bf{C}}_n},{\bf{C}}_n^E\} }\
&&\sum\limits_{j = 1}^J {{{\left( {{{\bf{C}}_n}} \right)}_j}} - \sum\limits_{j = 1}^J {{{\left( {{\bf{C}}_n^E} \right)}_j}}\label{eqn:sub4} \\
{\rm{s.t.}}&&{\left( {{{\bf{C}}_n}} \right)_j} - {\left( {{\bf{C}}_n^E} \right)_j} \ge {\left( {{\bf{C}}_{\min }^S} \right)_j}{\rm{ }}, \forall j, \\
&&{\rm{log}}\left( {\rho \left( {{{{\bf{\tilde B}}}_{nj}}diag\left( {{e^{{{\bf{C}}_n}}}} \right)} \right)} \right) \le 0{\rm{ }}, \forall j, \label{eqn:cons4}\\
&&{\rm{log}}\left( {\rho \left( {{\bf{\tilde B}}_{nj}^Ediag\left( {{e^{{\bf{C}}_n^E}}} \right)} \right)} \right) \le 0{\rm{ }}, \forall j,\label{eqn:cons5}
\end{eqnarray}
\emph{where $\rho \left( \cdot \right)$ is the the Perron-Frobenius eigenvalue of a nonnegative matrix}.\hfill\rule{8pt}{8pt}

\textbf{\emph{Proof:}} Please see Appendix C. \hfill\rule{8pt}{8pt}

Due to the log-convexity property of the Perron-Frobenius eigenvalue, the constraints \eqref{eqn:cons4} and \eqref{eqn:cons5} are convex. Now the concerned optimization
problem in \eqref{eqn:sub3} is a convex optimization problem and can be solved in polynomial time. However, such problem is very hard to solve by the traditional
sub-gradient convex optimization method because it is difficult to get the closed-form expressions of ${{{{\bf{C}}_n}}}$ and ${{\bf{C}}_n^E}$, which is essential
to achieve the solution. To deal with this challenge, a novel method which is effective for this problem is proposed in the following content.

The Lagrangian function of problem \eqref{eqn:sub4} can be expressed as:

\begin{align}
L({{\bf{C}}_n},{\bf{C}}_n^E,{\bm{\lambda }},{\bm{\beta }},{\bm{\mu }}) &= \sum\limits_{j = 1}^J {{{\left( {{{\bf{C}}_n}} \right)}_j}} - \sum\limits_{j = 1}^J {{{\left( {{\bf{C}}_n^E} \right)}_j}} \nonumber\\
&+ \sum\limits_{j = 1}^J {{\lambda _j}\left( {{{\left( {{{\bf{C}}_n}} \right)}_j} - {{\left( {{\bf{C}}_n^E} \right)}_j} - {{\left( {{{\bf{C}}_{{{\min }^S}}}} \right)}_j}} \right)} - \nonumber\\
&\sum\limits_{j = 1}^J {{\beta _j}{\rm{log}}\left( {\rho \left( {{{{\bf{\tilde B}}}_{nj}}diag\left( {{e^{{{\bf{C}}_n}}}} \right)} \right)} \right)} -\nonumber\\
&\sum\limits_{j = 1}^J {{\mu _j}{\rm{log}}\left( {\rho \left( {{\bf{\tilde B}}_{nj}^Ediag\left( {{e^{{\bf{C}}_n^E}}} \right)} \right)} \right)} ,\label{lagrian}
\end{align}
where ${\bm{\lambda }},{\bm{\beta }},{\bm{\mu }}$ are the Lagrange multipliers corresponding to the constraints.

Then, the Lagrangian dual function can be expressed as:
\begin{eqnarray}
g(\bm{\lambda} ,{\bm{\beta }},{\bm{\mu }} ) = \mathop {{\rm{max}}}\limits_{\{ {{\bf{C}}_n},{\bf{C}}_n^E\} } L({{\bf{C}}_n},{\bf{C}}_n^E,\bm{\lambda} ,{\bm{\beta }},{\bm{\mu }} ).\label{eqn:sub5}
\end{eqnarray}

Therefore, the dual optimization problem can be formulated as:
\begin{eqnarray}
\mathop {{\rm{min}}}\limits_{\{ \bm{\lambda} ,{\bm{\beta }},{\bm{\mu }}\} }
&&g(\bm{\lambda} ,{\bm{\beta }},{\bm{\mu }} ) \nonumber \\
{\rm{s.t.}}&&{\bm{\lambda }} \succeq 0,{\bm{\beta }} \succeq
0,\bm{\mu} \succeq 0. \label{eqn:sub6}
\end{eqnarray}

Since the optimization problem in \eqref{eqn:sub4} is convex, the duality gap between the primal problem and its dual problem is zero, which illustrates that the primal problem can be solved
by solving its dual problem\eqref{eqn:sub6}. The sub-gradient-based method can be utilized to solve \eqref{eqn:sub6} and the sub-gradient of the Lagrange multipliers in the dual function in the $i$-th iteration can be written as:
\begin{align}
&\nabla {\bm{\lambda} ^{(i + 1)}}(j) = {\left( {{\bf{C}}_n^{(i)}} \right)_j} - {\left( {{\bf{C}}_n^{E(i)}} \right)_j} - {\left( {{{\bf{C}}_{{{\min }^S}}}} \right)_j},1 \le j \le J,\\
&\nabla {\bm{\beta} ^{(i + 1)}}(j) = \log (\rho ({{\bf{\tilde B}}_{nj}}diag({e^{{\bf{C}}_n^{(i)}}}))),1 \le j \le J,\\
&\nabla {\bm{\mu} ^{(i + 1)}}(j) = \log (\rho ({\bf{\tilde B}}_{nj}^Ediag({e^{{\bf{C}}_n^{E(i)}}}))),1 \le j \le J
\end{align}
where ${\left( {{\bf{C}}_n^{(i)}} \right)_j}$ and ${\left( {{\bf{C}}_n^{E(i)}} \right)_j}$ are the capacity and leakage capacity of the $j$-th UE on the $n$-th subcarrier in the $i$-th iteration, respectively. $\nabla{\bm{\lambda} ^{(i + 1)}}(j)$, $\nabla{\bm{\beta} ^{(i + 1)}}(j)$ and $\nabla{\bm{\mu} ^{(i + 1)}}(j)$ denote the $j$-th sub-gradient corresponding to $\bm{\lambda}$, $\bm{\beta}$ and $\bm{\mu}$ utilized in the $(i+1)$-th iteration. Therefore, the update equations for the dual variables in the $(i+1)$-th iteration can be expressed as:
\begin{align}
&{\bm{\lambda} ^{(i+1)}}(j)=\nonumber\\
 & {\left[ {{\bm{\lambda} ^{(i)}}(j){\rm{ - }}{\xi _{{\bm{\lambda} ^{(i)}}}}(j)\left( {{{\left( {{\bf{C}}_n^{(i)}} \right)}_j} - {{\left( {{\bf{C}}_n^{E(i)}} \right)}_j} - {{\left( {{{\bf{C}}_{{{\min }^S}}}} \right)}_j}} \right)} \right]^ + }\\
&{\bm{\beta} ^{(i + 1)}}(j) = {\left[ {{\bm{\beta} ^{(i)}}(j){\rm{ - }}{\xi _{{\bm{\beta} ^{(i)}}}}(j)\log (\rho ({{{\bf{\tilde B}}}_{nj}}diag({e^{{\bf{C}}_n^{(i)}}})))} \right]^ + }\\
&{\bm{\mu} ^{(i+1)}}(j)={\left[ {{\bm{\mu} ^{(i)}}(j){\rm{ - }}{\xi _{{\bm{\mu} ^{(i)}}}}(j)\log (\rho ({\bf{\tilde B}}_{nj}^Ediag({e^{{\bf{C}}_n^{E(i)}}})))} \right]^ + }
\end{align}
where ${\xi _{{\bm{\lambda} ^{(i)}}}}(j)$, ${\xi _{{\bm{\beta} ^{(i)}}}}(j)$ and ${\xi _{{\bm{\mu} ^{(i)}}}}(j)$ are positive step sizes. According to the above derivations,
a sub-gradient method based iteration algorithm is presented to solve \eqref{eqn:sub6} as shown in \textbf{Algorithm 1}.

\begin{algorithm}[htb]
\renewcommand{\algorithmicrequire}{\textbf{Input:}}
\renewcommand\algorithmicensure {\textbf{Output:} }
\caption{Sub-gradient method based iteration algorithm for the outer loop optimziation}
\label{alg:Framwork}
\begin{algorithmic}[1]
\STATE Set the iteration index $i=1$. Initialize the Lagrange multiplier $\bm{\lambda}^{(i)} ,{\bm{\beta }}^{(i)},{\bm{\mu }}^{(i)}$, the step size $\bm{\xi} _{\bm{\lambda}},\bm{\xi} _{\bm{\beta}},\bm{\xi} _{\bm{\mu}}$, the maximum number of iterations ${I_{\max}}$ and the iteration threshold $\delta$.\\
\STATE \textbf{for} $1 \le i \le {I_{\max }}$ \\
\STATE ~~Calculate ${{{\bf{C}}_n}^{(i)},{\bf{C}}_n^{E(i)}}$ which can be expressed as
\begin{align}
&\left( {{{{\bf{C}}_n}^{(i)},{\bf{C}}_n^{E(i)}}} \right) = \nonumber\\
&\mathop {{\rm{argmin}}}\limits_{\{ {{{\bf{C}}_n}^{(i)},{\bf{C}}_n^{E(i)}}\} } L({{{\bf{C}}_n}^{(i)},{\bf{C}}_n^{E(i)}},\bm{\lambda}^{(i)} ,{\bm{\beta }}^{(i)},{\bm{\mu }}^{(i)} )\label{eqn:sub7}
\end{align}\\
\STATE ~~
\begin{align}
&{\bm{\lambda} ^{(i+1)}}(j) = \nonumber\\
&{\left[ {{\bm{\lambda} ^{(i)}}(j){\rm{ - }}{\xi _{{\bm{\lambda} ^{(i)}}}}(j)\left( {{{\left( {{\bf{C}}_n^{(i)}} \right)}_j} - {{\left( {{\bf{C}}_n^{E(i)}} \right)}_j} - {{\left( {{{\bf{C}}_{{{\min }^S}}}} \right)}_j}} \right)} \right]^ + }\nonumber
\end{align};\\
\STATE ~~${\bm{\beta} ^{(i + 1)}}(j) = {\left[ {{\bm{\beta} ^{(i)}}(j){\rm{ - }}{\xi _{{\bm{\beta} ^{(i)}}}}(j)\log (\rho ({{{\bf{\tilde B}}}_{nj}}diag({e^{{\bf{C}}_n^{(i)}}})))} \right]^ + }$;\\
\STATE ~~${\bm{\mu} ^{(i+1)}}(j)={\left[ {{\bm{\mu} ^{(i)}}(j){\rm{ - }}{\xi _{{\bm{\mu} ^{(i)}}}}(j)\log (\rho ({\bf{\tilde B}}_{nj}^Ediag({e^{{\bf{C}}_n^{E(i)}}})))} \right]^ + }$;\\
\STATE ~~\textbf{if} $\left| {{\bm{\lambda} ^{(i + 1)}}(j) - {\bm{\lambda} ^{(i)}}(j)} \right| + \left| {{\bm{\beta} ^{(i + 1)}}(j) - {\bm{\beta} ^{(i)}}(j)} \right| + \left| {{\bm{\mu} ^{(i + 1)}}(j) - {\bm{\mu} ^{(i)}}(j)} \right| \le \delta $
\STATE ~~~~break out;
\STATE ~~\textbf{end if}
\STATE ~~$i=i+1$;\\
\STATE \textbf{end for}
\STATE return ${{{\bf{C}}_n}^{(i)}-{\bf{C}}_n^{E(i)}}$ .
\end{algorithmic}
\end{algorithm}

To solve \eqref{eqn:sub7}, rewrite \eqref{lagrian} with given Lagrange multiplier $\bm{\lambda}^{(i)} ,{\bm{\beta }}^{(i)},{\bm{\mu }}^{(i)}$ as:
\begin{align}
L({{\bf{C}}_n},{\bf{C}}_n^E) &= \sum\limits_{j = 1}^J {{{\left( {{{\bf{C}}_n}} \right)}_j}} - \sum\limits_{j = 1}^J {{{\left( {{\bf{C}}_n^E} \right)}_j}}\nonumber\\
 &+ \sum\limits_{j = 1}^J {{\lambda _j^{(i)}}\left( {{{\left( {{{\bf{C}}_n}} \right)}_j} - {{\left( {{\bf{C}}_n^E} \right)}_j} - {{\left( {{{\bf{C}}_{{{\min }^S}}}} \right)}_j}} \right)} \nonumber\\
&-\sum\limits_{j = 1}^J {\beta _j^{(i)}{\rm{log}}\left( {\rho \left( {{{{\bf{\tilde B}}}_{nj}}diag\left( {{e^{{{\bf{C}}_n}}}} \right)} \right)} \right)} \nonumber\\
 &- \sum\limits_{j = 1}^J {\mu _j^{(i)}{\rm{log}}\left( {\rho \left( {{\bf{\tilde B}}_{nj}^Ediag\left( {{e^{{\bf{C}}_n^E}}} \right)} \right)} \right)} \\
 &= f\left( {{{\bf{C}}_n},{\bf{C}}_n^E} \right) + g\left( {{{\bf{C}}_n},{\bf{C}}_n^E} , \right)
\end{align}
where
\begin{align}
f\left( {{{\bf{C}}_n},{\bf{C}}_n^E} \right) = & - \sum\limits_{j = 1}^J {\beta _j^{(i)}{\rm{log}}\left( {\rho \left( {{{{\bf{\tilde B}}}_{nj}}diag\left( {{e^{{{\bf{C}}_n}}}} \right)} \right)} \right)} \nonumber\\
 &- \sum\limits_{j = 1}^J {\mu _j^{(i)}{\rm{log}}\left( {\rho \left( {{\bf{\tilde B}}_{nj}^Ediag\left( {{e^{{\bf{C}}_n^E}}} \right)} \right)} \right)} \nonumber\\
g\left( {{{\bf{C}}_n},{\bf{C}}_n^E} \right) = &\sum\limits_{j = 1}^J {{{\left( {{{\bf{C}}_n}} \right)}_j}} - \sum\limits_{j = 1}^J {{{\left( {{\bf{C}}_n^E} \right)}_j}} \\
&+ \sum\limits_{j = 1}^J {\lambda _j^{(i)}\left( {{{\left( {{{\bf{C}}_n}} \right)}_j} - {{\left( {{\bf{C}}_n^E} \right)}_j} - {{\left( {{{\bf{C}}_{{{\min }^S}}}} \right)}_j}} \right)} .
\end{align}

It is obvious that $g\left( {{{\bf{C}}_n},{\bf{C}}_n^E} \right)$ is closed proper convex function and $f\left( {{{\bf{C}}_n},{\bf{C}}_n^E} \right)$ is differentiable. So the proximal gradient method can be used to solve \eqref{eqn:sub7} according to the proximal theory. The proximal operator of $f\left( {{{\bf{C}}_n},{\bf{C}}_n^E} \right)$ is calculated as:
\begin{eqnarray}
{\rm{pro}}{{\rm{x}}_g}({{\bf{C}}_n},{\bf{C}}_n^E) = \left( {1 + \lambda _j^{(i)}} \right)\left( {{{\left( {{{\bf{C}}_n}} \right)}_j} - {{\left( {{\bf{C}}_n^E} \right)}_j} - {\bf{I}}} \right).
\end{eqnarray}

The partial derivatives of $f\left( {{{\bf{C}}_n},{\bf{C}}_n^E} \right)$ with respect to ${{\bf{C}}_n},{\bf{C}}_n^E$ are given by

\begin{eqnarray}
 \frac{{\partial f}}{{d{{\bf{C}}_n}}} = {\rm{ - }}\sum\limits_{j = 1}^J {\beta _j^{(i)}{\bf{x}}({{{\bf{\tilde B}}}_{nj}}diag({e^{{{\bf{C}}_n}}})) \circ {\bf{y}}({{{\bf{\tilde B}}}_{nj}}diag({e^{{{\bf{C}}_n}}}))} , \nonumber\\\\
 \frac{{\partial f}}{{d{\bf{C}}_n^E}} = {\rm{ - }}\sum\limits_{j = 1}^J {\mu _j^{(i)}{\bf{x}}({\bf{\tilde B}}_{nj}^Ediag({e^{{\bf{C}}_n^E}})) \circ {\bf{y}}({\bf{\tilde B}}_{nj}^Ediag({e^{{\bf{C}}_n^E}}))} .\nonumber\\
\end{eqnarray}
where ${\bf{x}}( \cdot )$ and ${\bf{y}}( \cdot )$ represent the right and left Perron-Frobenius eigenvalue, respectively.

It obvious that problem \eqref{eqn:sub7} is convex because it is a dual problem\textcolor[rgb]{1.00,0.00,0.00}{\cite{b8}}. Due to the proximal gradient method, the proximal operator of \eqref{eqn:sub7} has a fixed point, which is the optimal value of \eqref{eqn:sub7}, too.
Therefore, an iterative algorithm is proposed to find the fixed point as shown by the \textbf{Algorithm} \ref{alg:pro}. In fact, \textbf{Algorithm} 1
and \textbf{Algorithm} \ref{alg:pro} are the outer loop and inner loop, respectively. On each subcarrier, the \textbf{Algorithm} 1 is executed to solve the equivalent convex
optimization problem of D2D underlaying HetNets, while the \textbf{Algorithm} \ref{alg:pro} is utilized to solve the dual problem proposed in \textbf{Algorithm} 1.

The complexity of the proposed algorithm is analyzed and compared with the algorithm presented in \cite{r1} as follows.
\begin{itemize}
 \item The complexity of the subcarrier allocation scheme is linear with $O(N(H+LM+LK))$.
 \item The complexity of the power allocation algorithm is determined by Algorithm 1 and Algorithm 2. The complexity of Algorithm 1 is linear with the number of dual factors, i.e., $O(J)$. The complexity of Algorithm 2 is linear with the number of the elements of ${{{\bf{C}}_n}^{(i)},{\bf{C}}_n^{E(i)}}$, i.e., $O(J)$. Therefore, the complexity of the power allocation scheme is $O(J^2)$.
\end{itemize}

Therefore, the total complexity of the proposed algorithm is $O(NJ^2(H+LM+LK))$. The algorithm proposed in \cite{r1} has a complexity of $O(NK(N+M))$. Compared with the algorithm presented in \cite{r1}, the algorithm proposed in this paper considers both subcarrier and power allocation scheme at the expense of the complexity. Due to including subcarrier and power allocation executions, the proposed algorithm in this paper can be solved in polynomial time and is acceptable in practice.

\makeatletter
 \def\@eqnnum{{\normalfont \color{black} (\theequation)}}
 \makeatother

\begin{algorithm}[htb]
\renewcommand{\algorithmicrequire}{\textbf{Input:}}
\renewcommand\algorithmicensure {\textbf{Output:} }
\caption{The dual problem solution for the inner loop optimization}
\label{alg:pro}
\begin{algorithmic}[1]
\STATE Set the iteration index $s=1$. Initialize the Lagrange multiplier $\bm{\lambda}^{(i)} ,{\bm{\beta }}^{(i)},{\bm{\mu }}^{(i)}$ of the $i$-th iteration of \textbf{Algorithm} 1, the maximum number of iterations ${S_{\max}}$ and the iteration threshold $\eta$.\\
\STATE \textbf{for} $1 \le s \le {S_{\max }}$ \\
\STATE Calculate
\begin{eqnarray}
{\bf{C}}_n^{(s + 1)} &&= {\rm{pro}}{{\rm{x}}_g}({\bf{C}}_n^{(s)} - \frac{{\partial f}}{{d{\bf{C}}_n^{(s)}}},{\bf{C}}_n^{E(s)}),\\
{\bf{C}}_n^{E(s + 1)}&& = {\rm{pro}}{{\rm{x}}_g}({\bf{C}}_n^{(s)},{\bf{C}}_n^{E(s)} - \frac{{\partial f}}{{d{\bf{C}}_n^{E(s)}}}).
\end{eqnarray}\\
\STATE ~~\textbf{if} $\left| {{\bf{C}}_n^{(s + 1)} - {\bf{C}}_n^{(s)}} \right| + \left| {{\bf{C}}_n^{E(s + 1)} - {\bf{C}}_n^{E(s)}} \right| \le \eta $
\STATE ~~~~break out;
\STATE ~~\textbf{end if}
\STATE ~~$s=s+1$
\STATE \textbf{end for}
\STATE return ${\bf{C}}_n^{(s )}$ and ${\bf{C}}_n^{E(s)}$
\end{algorithmic}
\end{algorithm}

\section{Simulation Results and Discussions}

In this section, the secrecy capacity performance of the D2D communication underlaying HetNets and the corresponding optimization results are numerically evaluated with simulations. In our simulations, one HPN is concerned, and all LPNs are uniformly located in the circle with the distance of $1000 m$ whose center is the HPN. The cell radius of HPN and LPN are $800 m$ and $200 m$, respectively. The minimum allowed distance between HPN and users is $50$ m and the minimum allowed distance between LPN and users is $20$ m to protect users from radiation. It is assumed that $H = 2$, i.e., there are 2 HUEs {\color{red}who access to} the HPN. Denote $M = 5$ and $K = 5$, which suggests that 5 LUEs and 5 D2D links randomly locate in the coverage of each LPN.
All of LPNs share $N = 8$ subcarriers and each subcarrier occupies $200$ KHz. On each subcarrier, an eavesdropper exists in the concerned scenario and it locates randomly.

The total transmit power of HPN is 43 dBm and equally allocated on all subcarriers. The transmit power of the transmitter of DUE is $15$ dBm
and the distance between the transmitter and the receiver is $10$ m. It is assumed that the path loss model is expressed as $31.5 + 40.0*\log_{10}(d)$ for the D2D link and the LPN-to-LUE,
LPN-D2D and D2D-LUE link with short transmit distance, while $31.5 + 35.0*\log_{10}(d)$ is used for other long links, where $d$ denotes the distance between the transmitter and the receiver in meters.
The number of simulation snapshots is set at 1000. In all snapshots,
the fast-fading coefficients are all generated as independently and
identically distributed (i.i.d.) Rayleigh random variables with unit
variances.

\subsection{Convergence of the Proposed Algorithm}

\begin{figure}
\centering \vspace*{0pt}
\includegraphics[scale=0.56]{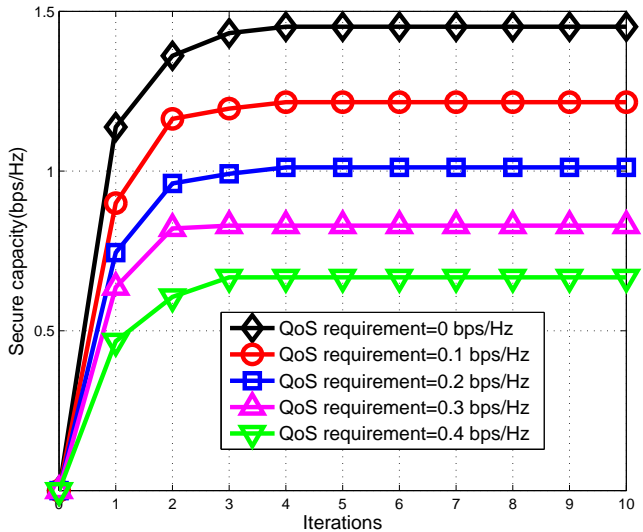}
\setlength{\belowcaptionskip}{-100pt} \caption{Convergence evolution
under different QoS requirements}\label{Converge}
\end{figure}

To evaluate the convergence performance of the algorithm under more users, we set $H=10$, $N=20$, $M=15$ and $K=15$, respectively. The convergence of the proposed algorithm with different QoS requirements is shown in Fig. \ref{Converge}. It can be obviously illustrated that the proposed algorithm converges within 5 iteration numbers for different QoS requirements, which suggests that the proposed algorithm can work efficiently. Besides, the different levels of QoS requirements have significant impacts on the performance of the secrecy capacity.
To evaluate
the influence of the QoS requirements, three QoS levels are set and the maximum allowed transmit power of LPNs is set at 24 dBm. Fig. \ref{Converge} shows that the secrecy capacity of the system decrease with the QoS requirement increasing.
In the case of high QoS requirement, more transmit power should be allocated to the user who is unable to achieve the QoS requirement, thus the system performance has a decrease. So there exists a balance between the system performance and the user performance.

\subsection{Secure Capacity Performances of the Proposed Solutions}

In this part, key factors impacting on the secrecy capacity performances of the proposed algorithm are evaluated. Since the transmit power constraint and the QoS requirement
are two main constraints in the proposed optimization problem, the impact of these two factors are evaluated in Fig. 3 and Fig. 4, respectively.
Besides, the average secrecy capacity per UE is more significant than the total secrecy capacity to evaluate the performance of the D2D underlaying HetNets when the number of UEs changes frequently.
Therefore, to evaluate the impact of the number of UEs to the D2D underlaying HetNets, the relationship between the number of UEs and the average secrecy capacity per UE should be evaluated.
Since LUEs usually contribute more capacity to the network than HUEs and DUEs, without loss of generality, only the relationship between the number of LUEs per LPN and the average secrecy capacity per LUE is evaluated in this part and the corresponding simulation result is shown in Fig. 5. It should be noted that in HetNets, HPNs are mainly deployed to provide seamless coverage and LPNs which are closer to users are deployed in hot spots to provide high capacity. Therefore, the capacity of the network is mainly determined by LPNs and only the maximum transmit power allowance of LPN is simulated in this paper.

In Fig. 3, the secrecy capacity performances under varied QoS requirements with a step size of 0.02 are compared among different maximum allowed transmit power.
When the QoS requirement is not sufficiently large, the secrecy performance decreases slowly with the increasing QoS requirement because most users can set a sufficiently high transmit power to
satisfy its QoS requirement. However, when the QoS requirement becomes large enough, more users can afford the QoS requirement and then more transmit power has to be allocated to these users to meet the QoS requirement. Furthermore, it can be illustrated from Fig. 3 that with the larger maximum allowed transmit power of LPN, the secrecy capacity performance increases, which has a similar trend with Fig. 4.

\begin{figure}
\centering \vspace*{0pt}
\includegraphics[scale=0.56]{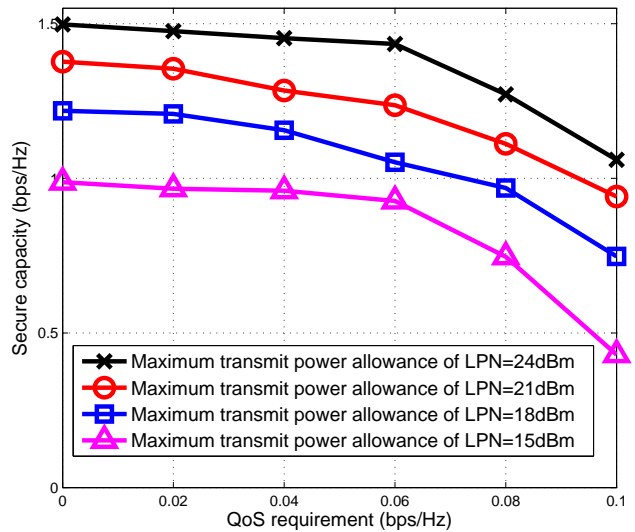}
\setlength{\belowcaptionskip}{-100pt} \caption{Secure capacity performance
comparisons under different QoS requirements}
\end{figure}

To further evaluate the impact of the maximum allowed transmit power
of LPNs on the secure capacity performances, four baselines are presented. The first baseline is the upper bound of this problem and it is calculated by the exhaustive method. We do the power allocation algorithm for each possible subcarrier allocation scheme and get the optimal subcarrier and power allocation for the problem. The second baseline is proposed by\textcolor[rgb]{1.00,0.00,0.00}{\cite{n3}}. In\textcolor[rgb]{1.00,0.00,0.00}{\cite{n3}}, users that cause severe interference are scheduled to the other subcarriers to alleviate interference. For example, if user A interferes other users seriously on the $i$-th subcarrier, it can be scheduled to the $j$-th subcarrier, on which the interference can be almost avoided. The third baseline is based on the classic orthogonal subcarrier allocation which has been widely used in HetNets. The fourth baseline is based on the fixed power scheme, i.e., the transmit power allocated to LUEs is same and fixed. Fig. 4 compares the secure capacity performance of different algorithms in terms of maximum allowed transmit power of LPNs. In this simulation case, the QoS requirement is set at 0 and the maximum allowed transmit power of LPNs varies within [14 dBm, 36 dBm] with the step size of 2 dBm. Fig. 4 shows that when the transmit power is not sufficiently high, the secure capacity performance increases with the transmit power rises because the interference is still not the bottleneck and the increment of secure capacity mainly depends on increasing transmit power. However, when the maximum allowed transmit power of LPNs is sufficiently high, the interference limits the increase of the secure capacity. Thus, no more transmit power is allocated to users. As shown in Fig. 4, the proposed algorithm is closed to the upper bound and outperforms the algorithm presented in\textcolor[rgb]{1.00,0.00,0.00}{\cite{n3}} because the reference\textcolor[rgb]{1.00,0.00,0.00}{\cite{n3}} just proposed some mechanisms to mitigate the influence of interference. And the algorithm presented in\textcolor[rgb]{1.00,0.00,0.00}{\cite{n3}} is better than the orthogonal subcarrier allocation due to the benefit of frequency reuse. Furthermore, the fixed power scheme has the worst performance because it takes no measures to optimize the secure capacity performance of the whole system.

\begin{figure}
\centering \vspace*{0pt}
\includegraphics[scale=0.5]{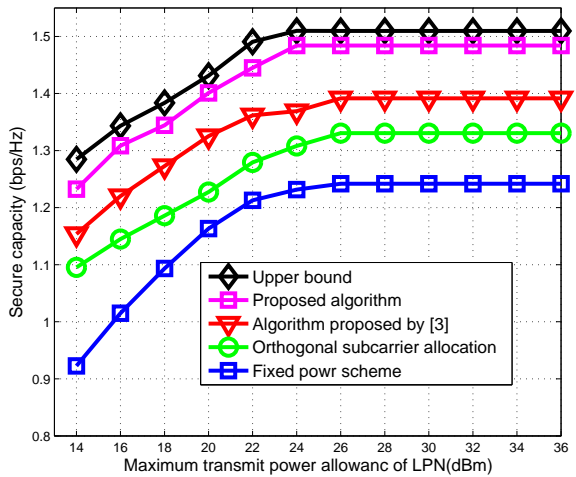}
\setlength{\belowcaptionskip}{-100pt} \caption{Secure capacity performance
comparisons under different maximum transmit power allowances of
LPNs}\label{txp}
\end{figure}

Fig. 5 illustrates the relationship between the number of LUEs and the average secrecy capacity per LUE to evaluate the impact of the number of UEs to the whole network.
In this simulation case, the QoS requirement is set at 0.1 bps/Hz and the maximum allowed transmit power of LPNs varies from 18 dBm to 22 dBm with a step size of 2 dBm.
The minimum number of LUEs per LPN is set to 1. Since the number of subcarriers is set to 8, which suggests that at most 8 LUEs can simultaneously connect to the LPN in each time slot in each LPN,
the maximum number of LUEs per LPN are set to 8. It can be seen in Fig. 5 that as the number of LUEs per LPN increases, the average secrecy capacity per LUE decreases because the intra-tier interference increases with more LUEs connecting to the same LPN.
The augment of intra-tier interference leads to the abatement of average secrecy capacity per LUE. However, it can be calculated from Fig. 5 that the total secrecy capacity of the whole LUEs keeps increasing as the number of LUEs per LPN rises, because when more LUEs connect to the LPN, more subcarriers can be allocated to these LUEs until the number of LUEs is equal to the number of the subcarriers.
Besides, Fig. 5 shows that the higher the maximum transmit power allowance of LPN is, the better the secrecy capacity of the whole network achieves, which has the same trend as shown in Fig. 3.

\begin{figure}
\centering \vspace*{0pt}
\includegraphics[scale=0.54]{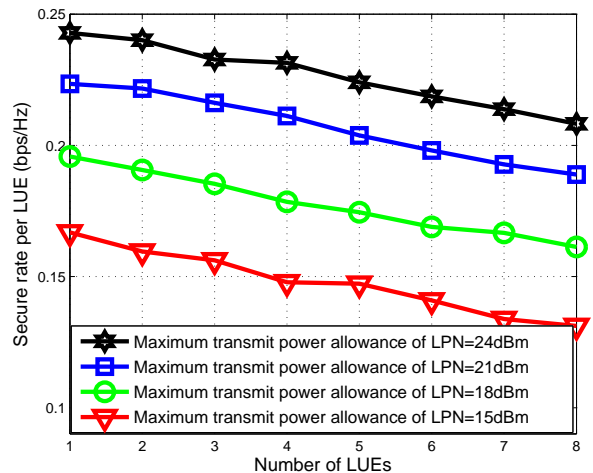}
\setlength{\belowcaptionskip}{-100pt} \caption{Average secure capacity per LUE comparisons under different number of LUEs per LPN}\label{txp}
\end{figure}

\section{Conclusion}

In this paper, the secrecy-optimized resource allocation for the device-to-device (D2D) communication underlaying heterogeneous networks (HetNets) has been researched.
In the concerned system model, there densely exist high power node and low power node with D2D communication, therefore, the inter-tier interference is always severe, which leads to the secrecy-optimized maximization problem being non-convex.
To solve this non-convex optimization problem, the primal non-convex optimization problem has been transformed into the convex issue with several steps.
Firstly, the objective function with several constraints are transformed into a matrix form. Secondly, the equivalent convex form of the maximization problem for the matrix form is derived
according to the Perron-Frobenius theory. Thirdly, the proximal operator of the transformed convex problem is evaluated. Then, a novel iterative algorithm based on the proximal theory
to solve the equivalent convex problem is proposed.

Simulation results have demonstrated that the secrecy capacity has a significant improvement and the proposed algorithm is effective and converges fast.
Furthermore, comparing with four baselines, the proposal has a significant performance gain on the secrecy capacity of the whole network. In the future, the non-ideal CSIs should be considered to optimize the secrecy capacity.
In addition, the advanced inter-tier interference in the physical layer and the dynamical queue characteristics in the upper layer should be jointly considered with the radio resource allocation to optimize the secrecy capacity in the D2D communication underlying HetNets.

\appendices

\section{Proof of \textbf{Theorem 1}}

Without loss of generality, let ${\bf{x}}_p$ denote the optimal ${\bf{x}}$. Thus, It can be derived as:
\begin{equation}\label{app2}
{{\bf{x}}_p} = \mathop {{\rm{argmax}}}\limits_{\{ {{\bf{x}}_i}\} } {\kern 1pt} {\kern 1pt} {\kern 1pt} {\kern 1pt} {\kern 1pt} \left( {f({{\bf{x}}_i},{\bf{y}}_i^*){\kern 1pt} } \right).
\end{equation}

Assume that there exists ${{\bf{x}}_j} \ne {{\bf{x}}^*}$ which is better than ${\bf{x}}_i$. Therefore, the following equation holds.
\begin{equation}
f({{\bf{x}}_p},{\bf{y}}_p^*) < f({{\bf{x}}_j},{\bf{y}}_j^*),
\end{equation}
which means that
\begin{equation}\label{app1}
{{\bf{x}}_j} = \mathop {{\rm{argmax}}}\limits_{\{ {{\bf{x}}_i}\} } {\kern 1pt} {\kern 1pt} {\kern 1pt} {\kern 1pt} {\kern 1pt} \left( {f({{\bf{x}}_i},{\bf{y}}_i^*){\kern 1pt} } \right).
\end{equation}

It is obviously that \eqref{app1}s conflict with \eqref{app2}. Thus ${\bf{x}}_p$ is the optimal $\bf{x}$ and theorem 1 holds.

\section{Proof of \textbf{Theorem 2}}

According to \eqref{append1}, it can be derived that ${e^{{{\left( {{{\bf{C}}_n}} \right)}_j}}} - 1 = {\left( {{{\bf{S}}_n}} \right)_j} = \frac{{{p_{nj}}}}{{{{\left( {{{\bf{F}}_n}{{\bf{p}}_n} + {{\bf{v}}_n}} \right)}_j}}}$ which is the same as:

\begin{equation}\label{append2}
\left( {{\rm{diag}}({e^{{{\bf{C}}_n}}}) - {\bf{I}}} \right)\left( {{{\bf{F}}_n}{{\bf{p}}_n} + {{\bf{v}}_n}} \right) = {{\bf{p}}_n}.
\end{equation}

Substituting \eqref{q} into \eqref{append2}, it can be derived as:
\begin{equation}\label{append3}
{\rm{diag}}({e^{{{\bf{C}}_n}}}){{\bf{q}}_n} = {{\bf{p}}_n} + {{\bf{q}}_n}.
\end{equation}

Multiply both sides of \eqref{append3} by ${{\bf{B}}_{nj}}$ which is given by \eqref{b}, we can have
\begin{align}
&\left( {{{\bf{F}}_n} + \frac{1}{{{{\left( {{{\bf{p}}_{\max }}} \right)}_j}}}{{\bf{v}}_n}\bm{\alpha} _j^{\rm{T}}} \right){\rm{diag}}({e^{{{\bf{C}}_n}}}){{\bf{q}}_n} \nonumber \\=
 &\left( {{{\bf{F}}_n} + \frac{1}{{{{\left( {{{\bf{p}}_{\max }}} \right)}_j}}}{{\bf{v}}_n}\bm{\alpha} _j^{\rm{T}}} \right){{\bf{p}}_n} + \left( {{{\bf{F}}_n} + \frac{1}{{{{\left( {{{\bf{p}}_{\max }}} \right)}_j}}}{{\bf{v}}_n}{\bm{\alpha}} _j^{\rm{T}}} \right){{\bf{q}}_n}\\
=&{{\bf{F}}_n}{{\bf{p}}_n} + \frac{1}{{{{\left( {{{\bf{p}}_{\max }}} \right)}_j}}}{{\bf{v}}_n}{\bm{\alpha }}_j^{\rm{T}}{{\bf{p}}_n} + \left( {{{\bf{F}}_n} + \frac{1}{{{{\left( {{{\bf{p}}_{\max }}} \right)}_j}}}{{\bf{v}}_n}{\bm{\alpha}} _j^{\rm{T}}} \right){{\bf{q}}_n}.
\end{align}

According to \eqref{a}, ${\bm{\alpha }}_j^{\rm T}{{\bf{p}}_n} \le {\left( {{{\bf{p}}_{\max }}} \right)_j}{\rm{ }}$. Thus
\begin{eqnarray}
\frac{1}{{{{\left( {{{\bf{p}}_{\max }}} \right)}_j}}}{{\bf{v}}_n}{\bm{\alpha }}_j^{\rm{T}}{{\bf{p}}_n} = {{\bf{v}}_n}\frac{{{\bm{\alpha }}_j^{\rm{T}}{{\bf{p}}_n}}}{{{{\left( {{{\bf{p}}_{\max }}} \right)}_j}}}\le {{\bf{v}}_n},
\end{eqnarray}
and
\begin{eqnarray}
&&{{\bf{F}}_n}{{\bf{p}}_n} + \frac{1}{{{{\left( {{{\bf{p}}_{\max }}} \right)}_j}}}{{\bf{v}}_n}{\bm{\alpha }}_j^{\rm{T}}{{\bf{p}}_n} + \left( {{{\bf{F}}_n} + \frac{1}{{{{\left( {{{\bf{p}}_{\max }}} \right)}_j}}}{{\bf{v}}_n}{\bm{\alpha}} _j^{\rm{T}}} \right){{\bf{q}}_n}\nonumber\\
 \label{append4}\le &&{{\bf{F}}_n}{{\bf{p}}_n} + {{\bf{v}}_n} + \left( {{{\bf{F}}_n} + \frac{1}{{{{\left( {{{\bf{p}}_{\max }}} \right)}_j}}}{{\bf{v}}_n}\alpha _j^{\rm{T}}} \right){{\bf{q}}_n}\\
 \label{append5}=&&{{\bf{q}}_n} + \left( {{{\bf{F}}_n} + \frac{1}{{{{\left( {{{\bf{p}}_{\max }}} \right)}_j}}}{{\bf{v}}_n}\alpha _j^{\rm{T}}} \right){{\bf{q}}_n}\\
 =&&\left( {{\bf{I}} + {{\bf{F}}_n} + \frac{1}{{{{\left( {{{\bf{p}}_{\max }}} \right)}_j}}}{{\bf{v}}_n}\alpha _j^{\rm{T}}} \right){{\bf{q}}_n},
\end{eqnarray}
where \eqref{append5} is derived by \eqref{q}. Now the formula can be derived as:
\begin{eqnarray}
&&\left( {{{\bf{F}}_n} + \frac{1}{{{{\left( {{{\bf{p}}_{\max }}} \right)}_j}}}{{\bf{v}}_n}\bm{\alpha} _j^{\rm{T}}} \right){\rm{diag}}({e^{{{\bf{C}}_n}}}){{\bf{q}}_n} \nonumber\\
&&\le \left( {{\bf{I}} + {{\bf{F}}_n} + \frac{1}{{{{\left( {{{\bf{p}}_{\max }}} \right)}_j}}}{{\bf{v}}_n}\alpha _j^{\rm{T}}} \right){{\bf{q}}_n}.
\end{eqnarray}

Substituting \eqref{b} into the above inequality, \eqref{bcons1} can be derived. Similarly, \eqref{bcons2} can be derived according to \eqref{be} and \eqref{qe}.

\section{Proof of \textbf{Theorem 3}}

Since ${\left( {{\bf{I}} + {{\bf{B}}_{nj}}} \right)^{ - 1}}$ exists, multiply both sides of constraint \eqref{eqn:cons2} by ${\left( {{\bf{I}} + {{\bf{B}}_{nj}}} \right)^{ - 1}}$, and the following equation can be derived as:
\begin{eqnarray}
&&{{\bf{B}}_{nj}}{\rm{diag}}({e^{{{\bf{C}}_n}}}){{\bf{q}}_n} \le \left( {{\bf{I}} + {{\bf{B}}_{nj}}} \right){{\bf{q}}_n}{\rm{ }}\nonumber \\
 \Leftrightarrow &&{\left( {{\bf{I}} + {{\bf{B}}_{nj}}} \right)^{ - 1}}{{\bf{B}}_{nj}}{\rm{diag}}({e^{{{\bf{C}}_n}}}){{\bf{q}}_n} \le {{\bf{q}}_n}\nonumber \\
 \Leftrightarrow && {{{\bf{\tilde B}}}_{nj}}{\rm{diag}}({e^{{{\bf{C}}_n}}}){{\bf{q}}_n} \le {{\bf{q}}_n}.
\end{eqnarray}

According to the subinvariance theorem\textcolor[rgb]{1.00,0.00,0.00}{\cite{b3}}, if a nonnegative matrix $\bf{A}$, a positive number $a$ and a nonnegative vector $\bf{v}$ satisfy ${\bf{Av}} \le a{\bf{v}}$,
then $\rho \left( {\bf{A}} \right) \le a$ and the equality holds if and only if ${\bf{Av}} = a{\bf{v}}$. Let ${\bf{A}} = {{{\bf{\tilde B}}}_{nj}}{\rm{diag}}({e^{{{\bf{C}}_n}}})$,
$a = 1$ and ${\bf{v}} = {{\bf{q}}_n}$ and rewrite \eqref{eqn:cons2} as ${\rho \left( {{{{\bf{\tilde B}}}_{nj}}diag\left( {{e^{{{\bf{C}}_n}}}} \right)} \right)} \le 1 $,
which is the same as ${\rm{log}}\left( {\rho \left( {{{{\bf{\tilde B}}}_{nj}}diag\left( {{e^{{{\bf{C}}_n}}}} \right)} \right)} \right) \le 0{\rm{ }}$.
Similarly, it can be derived that ${\rm{log}}\left( {\rho \left( {{\bf{\tilde B}}_{nj}^Ediag\left( {{e^{{\bf{C}}_n^E}}} \right)} \right)} \right) \le 0{\rm{ }}$.


\begin{thebibliography}{99}

\bibitem{n1}
V. Suryaprakash, J. Moller, and G. Fettweis, ``On the modeling and analysis of heterogeneous radio access networks using a poisson cluster process,'' \emph{IEEE Trans. Wireless Commun.}, vol. 14, no. 2, pp. 1035-1047, Feb. 2015.

\bibitem{n2}
J. Young, M. Hasna, and A. Ghrayeb, ``Modeling heterogeneous cellular networks interference using poisson cluster processes,'' \emph{IEEE
J. Sel. Areas Commun.}, vol. 33, no. 10, pp. 2182-2195, Oct. 2015.

\bibitem{n3}
S. Singh and J. Andrews, ``Joint resource partitioning and offloading in heterogeneous cellular networks,'' \emph{IEEE Trans. Wireless Commun.}, vol. 13, no. 2, pp. 888-901, Feb. 2014.

\bibitem{Hetnet_peng}
M. Peng, C. Wang, J. Li, H. Xiang, and V. Lau, ``Recent advances in underlay heterogeneous
networks: Interference control, resource allocation, and self-organization,'' \emph{IEEE Commun. Surveys \& Tutorials}, vol. 17, no. 2, pp. 700-729, second quarter, 2015.

\bibitem{review_add}
M. Jo, T. Maksymyuk, B. Strykhalyuk, and C.H. Cho, "Device-to-device-based heterogeneous radio access network architecture for mobile cloud computing," \emph{IEEE Wireless Commun.}, vol. 22, no. 3, pp. 50-58, Jun. 2015.

\bibitem{1}
K. Doppler \emph{et. al.}, ``Device-to-device communication as an underlay to LTE-advanced networks," \emph{IEEE Communications Magazine}, vol. 47, no. 12,
pp. 42--49, Dec. 2009.

\bibitem{n4}
H. Dong, W. Kae, S. Wha, and G. Dong, ``Two-stage semi-distributed resource management for Device-to-Device communication in cellular networks,'' \emph{IEEE Trans. Wireless Commun.}, vol. 13, no. 4, pp. 1908-1920, Apr. 2014.

\bibitem{2}
C. Yu, K. Doppler, C. B. Ribeiro, and O. Tirkkonen, ``Resource allocation optimization for device-to-device communication underlaying cellularn networks," \emph{IEEE Trans. Wireless Commun.}, vol. 10, no. 8,
pp. 2752--2763, Aug. 2011.

\bibitem{3}
T. Peng, Q. Lu, H. Wang, S. Xu, and W. Wang, ``Interference avoidance mechanisms in the hybrid cellular and device-to-device systems," in \emph{Proc. 2009 IEEE PIMRC}, pp. 617--621

\bibitem{4}
H. Min, J. Lee, S. Park, and D. Hong, ``Capacity enhancement using an interference limited area for device-to-device uplink underlaying cellular networks," \emph{IEEE Trans. Wireless Commun.}, vol. 10, no. 12,
pp. 3995--400, Dec. 2011.

\bibitem{5}
X. Chen, L. Chen, M. Zeng, X. Zhang, and D. Yang, ``Downlink resource allocation for device-to-device communication underlaying cellular networks," in \emph{Proc. Pers. Indoor, Mobile Radio Commun. Conf.}, pp. 232--237

\bibitem{6}
F. Li, M. Lei, and F. Gao, ``Device-to-device (D2D) communication in MU-MIMO cellular networks," in \emph{Proc. Global Commun. Conf..}, pp. 3538--3587, Dec. 2012.

\bibitem{r1}
B. Peng, C. Hu, T. Peng and W. Wang, ``Optimal resource allocation for multi-d2d links underlying OFDMA-based communications," in \emph{Proc. International Conference on Wireless Communications, Networking and Mobile Computing (WiCOM)}, pp. 1--4, Shanghai, 2012.

\bibitem{r2}
W. Liu, Y. Yang, T Peng, and W. Wang, "Optimal resource allocation scheme for
satisfying the data rate requirement in hybrid network of D2D-cellular," \emph{Journal of Computers}, vol. 9, no. 5, may. 2014.

\bibitem{n6}
X. Tang, R. Liu, P. Spasojevic, and H. Poor, ``Interference assisted secret communication,'' \emph{IEEE Trans. Inf. Theory}, vol. 57, no. 5, pp. 3153-3167, May 2011.

\bibitem{n7}
L. Dong, Z. Han, A. Petropulu, and H. Poor, ``Improving wireless physical layer security via cooperating relays,'' \emph{IEEE Trans. Signal Process.}, vol. 58, no. 3, pp. 1875-1888, Mar. 2010.

\bibitem{n8}
Y. Liang, H. Poor, and S. Shamai, ``Secure communication over fading channels,'' \emph{IEEE Trans. Inf. Theory}, vol. 54, no. 6, pp. 2470-2492, Jun. 2008.

\bibitem{7}
X. Tang, R. Liu, P. Spasojevi¡äc, and H. Poor, ``Interference assisted secret communication," \emph{IEEE Trans. Inf. Theory}, vol. 57, no. 5, pp. 3153--3167, May 2011.

\bibitem{8}
Z. Chu, K. Cumanan, M. Xu, and Z. Ding, ``Robust secrecy rate optimisations for multiuser multiple-input-single-output channel with device-to-device communications," \emph{IET Communications}, vol. 9, no. 3, pp. 396--403, Feb. 2015.

\bibitem{9}
H. Zhang, T. Wang, L. Song, and Z. Han, ``Radio resource allocation for physical-layer security in D2D underlay communications," in \emph{Proc. 2014 IEEE ICC}, pp. 2319--2324, June. 2014.

\bibitem{n5}
J. Yue, C. Ma, H. Yu, and W. Zhou, ``Secrecy-Based Access Control for Device-to-Device Communication Underlaying Cellular Networks," \emph{IEEE Commun. Lett.}, vol. 17, no. 11, pp. 2068-2071, Nov. 2013.




\bibitem{10}
V. Blondel, L. Ninove, and P. Dooren, ¡°An affine eigenvalue problem on the nonnegative orthant,¡± \emph{Linear Algebra Its Appl}., vol. 404, pp. 69¨C84, 2005.

\bibitem{11}
U. Krause, ¡°Concave Perron¨CFrobenius theory and applications,¡± \emph{Nonlinear Anal}, vol. 47, no. 2001, pp. 1457¨C1466, 2001.

\bibitem{12}
C. Tan, S. Friedland, and S. Low, ¡°Nonnegative matrix inequalities and their application to nonconvex power control optimization,¡± \emph{SIAM J. Matrix Anal. Appl}., vol. 32, no. 3, pp. 1030¨C1055, 2011.

\bibitem{31}
A. Tolli, H. Pennanen, and P. Komulainen, "Decentralized minimum power multi-cell beamforming with limited backhaul signaling," \emph{IEEE Transactions on Wireless Commun.}, vol. 10, no. 2, pp. 570-580, Feb. 2011.

\bibitem{32}
W. Xu and X. Wang, "Pricing-based distributed downlink beamforming in multi-cell OFDMA networks," \emph{IEEE J. Sel. Areas Commun.}, vol. 30, no. 9, pp. 1605-1613, Oct. 2012.

\bibitem{41}
B. Dai and W. Yu, "Sparse beamforming and user-centric clustering for downlink cloud radio access network," \emph{IEEE Access}, vol. 2, pp. 1326-1339, Oct. 2014.
\bibitem{42}
D. W. K. Ng, E. S. Lo, and R. Schober, "Energy-efficient resource allocation in multi-cell OFDMA systems with limited backhaul capacity," \emph{IEEE Transactions on Wireless Commun.}, vol. 11, no. 10, pp. 3618-3631, Oct. 2012.
\bibitem{43}
C. Xiong, G. Y. Li, S. Zhang, Y. Chen and S. Xu, "Energy- and spectral-efficiency tradeoff in downlink OFDMA networks," \emph{IEEE Transactions on Wireless Commun.}, vol. 10, no. 11, pp. 3874-3886, Nov. 2011.

\bibitem{111}
C. Xiong, G.Y. Li, S. Zhang, Y. Chen, and S. Xu, ``Energy-efficient resource allocation in ofdma networks," \emph{IEEE Trans. Commun.}, vol. 60, no. 12, pp. 3767-3778, Dec. 2012.

\bibitem{222}
X. Xiao, X. Tao, and J. Lu, ``QoS-aware energy-efficient radio resource scheduling in multi-user ofdma systems," \emph{IEEE Commun. Lett.}, vol. 17, no. 1, pp. 75-78, Jan. 2013.

\bibitem{333}
W. Dang, M. Tao, H. Mu, and J. Huang, ``Subcarrier-pair based resource allocation for cooperative multi-relay OFDM systems," \emph{IEEE Trans. Wireless Commun.}, vol. 9, no. 5, pp. 1640-1649, May 2010.

\bibitem{detected}
J. Zhang, L. Fu, and X. Wang, ``Asymptotic analysis on secrecy capacity in large-scale wireless networks," \emph{IEEE/ACM Trans. Netw.}, vol. 22, no. 1, pp. 66-79, Feb. 2014.

\bibitem{fullCSI1}
A. Khisti and G. W. Wornell, ``Secure transmission with multiple antennas I: The MISOME wiretap channel," \emph{IEEE Trans. Inf. Theory}, vol. 56, no. 7, pp. 3088-3104, Jul. 2010.

\bibitem{fullCSI2}
F. Oggier and B. Hassibi, ``The secrecy capacity of the MIMO wiretap channel," \emph{IEEE Trans. Inf. Theory}, vol. 57, no. 8, pp. 4961-4972, Aug. 2011.

\bibitem{fullCSI3}
T. Liu and S. Shamai, ``A note on the secrecy capacity of the multiple-antenna wiretap channel," \emph{IEEE Trans. Inf. Theory}, vol. 55, no. 6, pp. 2547-2553, Jun. 2009.

\bibitem{NP1}
D. W. K. Ng, E. S. Lo, and R. Schober, ``Energy-efficient resource allocation in multi-cell OFDMA systems with limited backhaul capacity," \emph{IEEE Transactions on Wireless Commun.}, vol. 11, no. 10, pp. 3618-3631, Oct. 2012.

\bibitem{NP2}
G. Lim, C. Xiong, L. J. Cimini, and G. Y. Li, ``Energy-efficient resource allocation for OFDMA-based multi-RAT networks," \emph{IEEE Transactions on Wireless Commun.}, vol. 13, no. 5, pp. 2696-2705, May 2014.

\bibitem{NP3}
M. Tao, Y. C. Liang and F. Zhang, ``Resource allocation for delay differentiated traffic in multiuser OFDM systems," \emph{IEEE Transactions on Wireless Commun.}, vol. 7, no. 6, pp. 2190-2201, Jun. 2008.

\bibitem{b8}
S. Boyd and L. Vanderberghe, \emph{Convex Optimization}, Cambridge University Press, 2004.

\bibitem{b3}
C. Tan, S. Friedland, and S. Low, ``Nonnegative matrix inequalities
and their application to nonconvex power control optimization,'' \emph{SIAM
J. Matrix Analysis Appl}, vol. 32, no. 3, pp. 1030-1055, 2011.





\end{thebibliography}
\end{document}